\pdfoutput=1

\documentclass[10pt,a4paper]{article}

\usepackage{lipsum}
\usepackage{tocbibind}
\usepackage[bookmarksopen=false,breaklinks,colorlinks,linkcolor=blue,citecolor=blue,urlcolor=blue]{hyperref}
\usepackage{bookmark}

\usepackage{cite}
\usepackage[margin=1in]{geometry}
\usepackage{caption}
\usepackage{authblk}
\usepackage{amssymb,graphicx,multirow,multicol,booktabs,color}
\usepackage{float}
\usepackage{bm}

\usepackage{epstopdf}  

\usepackage{amsmath}
\numberwithin{equation}{section}

\usepackage[noend]{algpseudocode} 
\usepackage{algorithmicx,algorithm}

\usepackage{threeparttable}

\renewcommand{\thefigure}{\arabic{figure}}

\usepackage{subfigure}

\begin{document} 
	
	\title{\textbf{Local False Discovery Rate Estimation with Competition-Based Procedures for Variable Selection}}
	\author[1,2]{Xiaoya Sun}
	\author[1,2]{Yan Fu\thanks{ Correspondence: yfu@amss.ac.cn}}
	\small
	\affil[1]{\small CEMS, NCMIS, RCSDS, Academy of Mathematics  and Systems Science, Chinese Academy of Sciences, Beijing 100190, China.}
	\small
	\affil[2]{\small School of Mathematical Sciences,	University of Chinese Academy of Sciences,		Beijing 100049, China.}

	\date{} 
\maketitle

\vspace{-20pt}
\begin{abstract}
		Multiple hypothesis testing has been widely applied to problems dealing with high-dimensional data, e.g., selecting significant variables and controlling the selection error rate.
	The most prevailing measure of error rate used in the multiple hypothesis testing is the false discovery rate (FDR). 
	In recent years, local false discovery rate (fdr) has drawn much attention, due to its advantage of accessing the confidence of individual hypothesis. 
	However, most methods estimate fdr through $p$-values or statistics with known null distributions, which are sometimes not available or reliable. 
	Adopting the innovative methodology of competition-based procedures, e.g., knockoff filter, this paper proposes a new approach, named TDfdr, to local false discovery rate estimation, which is free of the $p$-values or known null distributions. 
	Simulation results demonstrate that TDfdr can accurately estimate the fdr with two competition-based procedures.
	In real data analysis, the power of TDfdr on variable selection is verified on two biological datasets.

\end{abstract}

\paragraph{Keywords}
	Multiple hypothesis testing; Variable selection; Local false discovery rate; Target-decoy; Knockoff; Null proportion estimation
	
\numberwithin{equation}{section}

\section{Introduction}

Multiple hypothesis testing is widely used in fields where large-scale data are produced, such as genomics, proteomics and massive
social science surveys. 
The aim of multiple hypothesis testing is to make assertions simultaneously for many hypotheses while controlling a certain type of error rate. 
Formally, suppose one is interested in simultaneously testing $m$ hypotheses, $H_1, H_2, \ldots, H_m$, with $H_j=0$ representing that the null hypothesis holds and $H_j=1$ otherwise. The goal of multiple hypothesis testing is rejecting a set of hypotheses $\{H_j | j \in R,~j=1,\ldots,m\}$, with an error rate controlled under a predetermined threshold $q \in (0,1)$ (here the set $R$ denotes the index set of the rejected hypotheses).
Specifically, in the cases where the hypotheses are about the significances of variables, the problem of multiple hypothesis testing is equivalent to variable selection, which plays an important role in high-dimensional data analysis. 

False discovery rate (FDR) has become the prevailing error rate used in multiple hypothesis testing since this concept was proposed by 
Benjamini and Hochberg\cite{BH:1995}.
FDR is defined as 
	$$FDR = 
	E\left\{ \frac{\#\{H_j: H_j =0, ~j \in R\}}{\#\{H_j : j \in R\} \vee 1 } \right\}.$$
FDR measures the expected proportion of falsely rejected hypotheses among all the rejected ones. 
It overcomes the conservativeness of previous measures such as PFER (per-family error rate) and FWER (familywise error rate), and thus can enhance power. 

Since the first procedure for FDR control was proposed
by 
Benjamini and Hochberg 
\cite{BH:1995} (shorted as "BH" hereafter), a variety of procedures have 
emerged. 
Amendments to the BH procedure were made to remove the independence assumption\cite{BY:2001,BY:2006}. Storey \cite{Storey:2002} developed the Bayesian framework of FDR, leading to a direct control of FDR. After that, the  concept of positive FDR was proposed \cite{Storey:2003}, along with a new quantity called $q$-value, which gives a Bayesian interpretation of FDR. 

While different procedures were created to control FDR, most if not all of them use $p$-value as the significance measure of tests. In situations where $p$-values are unavailable and general statistics or scores are produced, for instance, by machine learning methods, these classic procedures are inapplicable. 
Even though $p$-values are provided or produced in some way
, the potential flaws of $p$-value such as its inaccuracy in measuring significance
or 
sensitivity to sample size 
could make the FDR out of control.

In recent years, a new class of FDR control procedures have been proposed and  
become increasingly popular, which rely on the competition between the original variables and their "fake" counterparts, and do not require $p$-values or statistics with known null distributions \cite{knockoff:2015, He:2018, He:2018new, emery:2021, deeppink}. 
The fake variables resemble the original variables in statistical characteristics, but are irrelevant to the response. 
As far as we know, the idea of using  competition for FDR control originates from the target-decoy (TD) search strategy \cite{Elias:2007} used in the mass spectrometry-based proteomics, which specifically takes the advantage of competitive decoy peptide sequences to estimate and control the FDR of peptide identifications. He et al. set up the theoretical foundation of the TD search strategy \cite{He:2013, He:2015}.
Later, they extended the TD approach to the two-group study from a general perspective by introducing the decoy permutations of original samples. Their approach, named TDFDR, achieves FDR control for independent variables \cite{He:2018, He:2018new}.

Knockoff filter is a better-known competition-based method that also uses fake variables, called knockoffs, to control FDR \cite{knockoff:2015}. It has raised an enormous stream of FDR researches using knockoffs.
Originally aiming at variable selection in the linear regression problem, the knockoff filter achieves FDR control for correlated variables in specific settings by constructing the knockoffs in an elaborated manner. 
Afterwards, a series of methods were proposed to relax the restrictions on  the original model setting, such as the sample size and the linearity \cite{knockoff:2017panning,knockoff:2019robust,knockoff:2019casecontrol,knockoff:2017pfilter}, thus allowing the knockoff method to fit more complex situations. 

Both TDFDR and knockoff filter divide variables into two groups through a competition procedure.
Then they use the number of the "fake" variables to estimate that of the falsely rejected true nulls, and further, achieve the control of FDR. 
Details of the two methods are given in Supplementary S1.

While FDR has been dominantly used as a most powerful measure in multiple hypothesis testing, 
it measures the global error rate of a set of hypotheses rather than individual hypotheses. 
Efron et al.\cite{Efron:2001} proposed the concept of the local false discovery rate (fdr) from a new perspective. For a specific score $x$, the fdr at it is defined as
$$	fdr(x) = \frac{\pi_0 f_0(x)}{f(x)} = \frac{\pi_0 f_0(x)}{\pi_0 f_0(x) + \pi_1 f_1(x)},$$
where $\pi_0$ and $\pi_1$ are the proportions of true and false null hypotheses respectively, $f_0$ and $f_1$ are the density functions of scores corresponding to the true and false null hypotheses, and $f(x)$ is the density of the mixture distribution.
Defined in this way, 
fdr is able to measure the error rates of individual hypotheses \cite{Efron:2001, Efron:2002}. 
Meanwhile, the problem becomes the "estimation" of fdr rather than the "control" of FDR.
In turn, FDR can be derived from fdr using the relationship between them \cite{Efron:2002}
\begin{equation}\label{fdr2FDR}
	\operatorname{FDR}(x) 
	=E_{f}\{\operatorname{fdr}(X) \mid X \leq x\},
\end{equation}
where $E_f$ represents expectations with respect to the mixture density $f(x)$.

Many approaches were designed to estimate fdr, such as \cite{Pan:2003, Pounds:2003, Allison:2002, Broberg:2004, Aubert:2004, Liao:2004, Do:2005}. 
While all these methods have their respective innovations and effectiveness,
a problem with them is the use of $p$-values. First, with the opinion that $p$-value may give 
ambiguous information on a hypothesis, especially in the 
cases of high dimensional data in multiple hypothesis testing, these methods cannot provide sufficient reliability in fdr estimation. Second, in the cases where $p$-values are not accessible, these methods would be inapplicable.

When the inputs are not restricted to $p$-values, several approaches arise with different assumptions. 
Efron and Tibshirani \cite{Efron:2002} proposed an empirical Bayes approach, named locfdr, which employs maximum likelihood and the optional central matching method to estimate fdr.
Robin et al.\cite{Robin:2007} and Guedj et al.\cite{Robin:2009} proposed a semi-parametric approach for fdr estimation called "kerfdr". Under the assumption that the null distribution is known, kerfdr estimates the non-null distribution and fdr iteratively. 
Jeong et al.\cite{Jeong:2020} proposed a semi-parametric mixture method for fdr estimation, using Efron's methodology of empirical null and log-concave density estimation for the non-null distribution. 
Bickel and Rahal\cite{Bickel:2019} proposed the CFDR method to estimate fdr through transforming an estimated FDR.
The methods above are either parametric or semi-parametric, relying on the corresponding assumptions for different inputs. This introduces the risk of misspecification of the prior distribution. Parametric methods 
might sometimes fail or lead to inferior performance when the model assumptions are not met. As the generating mechanism of data in practice become increasingly complex, the limitations of parametric methods would become more apparent. As far as we know, the use of competitive fake variables has not yet been explored for  fdr estimation free of $p$-values.

In approaches to FDR and fdr, the proportion of true null hypotheses ($\pi_0$) plays an important role. The original BH procedure treats $\pi_0$ as unknown and allows it to be as large as one, which would seriously decrease the power if the real value of $\pi_0$ is small. Many approaches have been proposed to estimate $\pi_0$, most of which are also based on the $p$-value. 
Storey\cite{Storey:2002} estimated  $\pi_0$ using the 
property
that the null $p$-values follow $U(0,1)$ distribution. 
Langaas et al. \cite{Langaas:2005} estimated $\pi_0$ based on nonparametric maximum likelihood estimation of the $p$-value density.
	Instead of $p$-values, Efron\cite{Efron:2007} estimate $\pi_0$ and $f_0$ simultaneously using $z$-values through several optional methods, whereas the methods rely on the normality  assumption of $f_0$.
More accurate estimation of $\pi_0$ is still a major goal of current FDR researches \cite{Neumann:2021, Biswas:2022}.

In this paper, we proposed a new method for estimating fdr, which is called  TDfdr, relying on the competition-based procedures. This method does not lean on the $p$-value and can handle general scores or test statistics with or without known distributions. 
By treating $\pi_0$ as a special form of FDR, we first exploit the competition procedure to obtain an estimator of $\pi_0$. 
Then, we estimate the null distribution $f_0$ from the competitive "fake" variables through the kernel density method.
Finally, 
we utilize the framework of kerfdr 
\cite{Robin:2007, Robin:2009} 
to estimate the non-null distribution $f_1$ and fdr simultaneously.
Simulation studies demonstrate that TDfdr can accurately estimate the fdr for the TDFDR and knockoff competition procedures. The power of TDfdr is also investigated on two real biology datasets.

The remainder of the paper is arranged as follows.
Section 2 describes the estimation procedure and implementation algorithm of TDfdr.
Results of simulation comparison and real data analysis are given in Section 3 and 4,  respectively. Section 5 concludes the paper. 


\section{Local false discovery rate estimation}
	\label{sec:localfdr}
	
	TDfdr is built on the competition-based procedures, a new class of methods that apply to general scores other than $p$-values. To better illustrate the TDfdr method, we introduced two competition-based procedures, TDFDR\cite{He:2018, He:2018new} and knockoff filter\cite{knockoff:2015}, in Supplementary S1 with details. 
	Here, we continue to use the notations of TDFDR to describe TDfdr, although the framework of knockoff filter can be applied similarly to TDfdr. 
	
	Assume that we have $m$ simultaneous tests, each of which has the null hypothesis $H_j,~j=1,\ldots,m$. 
	For simplicity, we write $H_{j}=0$ when the null hypothesis $H_{j}$ holds and $H_{j}=1$ otherwise. 
	In the situation of variable selection, the purpose of tests becomes judgment of the significances of variables, and we will describe our TDfdr method in terms of variable selection as in the TDFDR and knockoff methods. 
	TDfdr aims to estimate the fdr of individual variables.

	As in the labelling step of TDFDR method described in Supplementary S1, the variables are separated into the target ($\mathcal{T}$) and decoy ($\mathcal{D}$) groups, which are defined as $\mathcal{T}:= \left\{j=1,2, \ldots, m : L_{j}=T\right\}$ and $\mathcal{D}:= \left\{j=1,2, \ldots, m : L_{j}=D\right\}$, respectively, where $L_j$ is the label of variable $j$ obtained from the competition procedure. The final scores, $S_j$'s, of the variables in the two groups are called  "target scores" and "decoy scores", respectively. 
	Further, we divide the variables into several subsets according to their labels and significances. 
	Define $\mathcal{T}_0 := \left\{j=1,2, \ldots, m : L_{j}=T, ~H_j=0\right\}$, $\mathcal{T}_1 := \left\{j=1,2, \ldots, m : L_{j}=T, ~H_j=1\right\}$. Subsets $\mathcal{D}_0$ and $\mathcal{D}_1$ are similarly defined with the label replaced by $L_j=D$.
	Note that only $\mathcal{T}$ and $\mathcal{D}$ are observable, and the four sets $\mathcal{T}_0$, $\mathcal{T}_1$, $\mathcal{D}_0$, $\mathcal{D}_1$ are unobservable in practice.

	Now we introduce three assumptions of TDfdr.
	
	\noindent Assumption 1: 
	The probabilities of a null variable being  labelled as target or decoy are  equal.
	
	\noindent Assumption 2: The final scores of decoy variables and null target variables have the same probability distribution. 
	
	\noindent Assumption 3: The probability of a non-null variable being  labelled as decoy is ignorable.

	Note that Assumptions 1 and 2 are the basis of all competition-based FDR control procedures and are naturally satisfied. 
	As for Assumption 3, we have good reasons to believe it holds widely. Can we consider a variable to be significant if it cannot win the competition with its fake insignificant counterpart?
	
	With these assumptions, TDfdr only pays attention to the target variables, and leaves the decoy ones as true nulls (insignificant variables) directly. 
	Thus, the aim of TDfdr becomes estimation of the fdr of target variables, which we call "target fdr", 
	\begin{equation}
		fdr_t(S_j) = \frac{\pi_{0t} f_{0t}(S_j)}{\pi_{0t} f_{0t}(S_j) + \pi_{1t} f_{1t}(S_j)}
		, \quad j \in \mathcal{T}
		\label{equ:realfdrt}
	\end{equation}
	where $\pi_{0t}$ represents the proportion of true nulls in target variables and $\pi_{1t} = 1-\pi_{0t}$,  $f_{0t}$ is the density function of scores of the null target variables, and $f_{1t}$ is the density of scores of the non-null target variables.

	Regarding the unknown quantities in Equation \ref{equ:realfdrt}, TDfdr first uses the decoy scores to estimate $\pi_{0t}$, as well as $f_{0t}$. With these two quantities, TDfdr then adopts the framework of a  semi-parametric method \cite{Robin:2007, Robin:2009} 
	to iterate $f_{1t}$. In the iteration procedure, the desired target fdr values can be calculated simultaneously.

	\subsection{$\pi_{0t}$ estimation}
	
	TDfdr employs the 
	competition procedure to estimate $\pi_{0t}$. The real value of $\pi_{0t}$ is
	\begin{equation}
		\pi_{0 t}
		=\frac{|\mathcal{T}_0|}{|\mathcal{T}|}
		={\frac{\#\left\{j=1,2, \ldots, m : L_{j}=T, ~H_j=0\right\}}{\#\left\{j=1,2, \ldots, m : L_{j}=T\right\}}}
		\label{equ:realpi0t}
	\end{equation}
	where $\mathcal{T}_0
	$ is unknown and is of our interest.

	Relying on the assumptions, we are able to estimate the number of null target variables $|\mathcal{T}_0|$ using the number of decoy variables $|\mathcal{D}|$.
	So we can estimate $\pi_{0t}$ as 
		$$\widehat{\pi_{0 t}}=\frac{|\mathcal{D}|}{|\mathcal{T}|}={\frac{\#\left\{j=1,2, \ldots, m : L_{j}=D \right\}}{\#\left\{j=1,2, \ldots, m : L_{j}=T\right\}}}$$
	If the estimated $\pi_{0t}$ is larger than 1, it is set  as 1, i.e.,
		$	\widehat{\pi_{0 t}}=\min \left\{\widehat{\pi_{0 t}}, 1\right\}
		\label{equ:regular}  $.
	
	Besides, $\pi_0$ can also be estimated using the similar idea,
	\begin{equation}
		\widehat{\pi_{0}} = \frac{2 |\mathcal{D}|}{m}= 
		\frac{2\#\left\{j=1,2, \ldots, m : L_{j}=D, ~H_j =0\right\}}
		{m}
		\label{equ:pi0}
	\end{equation}
	Note that if the probabilities in Assumption 1 are not equal, as long as they are constants, say $r$ for decoy and $1-r$ for target, we can estimate the $|\mathcal{T}_0|$ as $\frac{1-r}{r}|\mathcal{D}|$.

	\subsection{$f_{0t}$ estimation}
	
	The second part is the estimation of $f_{0t}$, i.e., the probability density function of
	the scores of true null target variables. Note that the Assumption 3 aforementioned 
	illustrates that there is an extremely small probability that a  non-null variable is labelled as decoy, which combining Assumption 1 means that 
	among all variables, the probabilities of a variable being decoy and that of a variable being null target, are equal.
	Therefore, the decoy variables are good resources to approximately describe the behavior of null target variables.
	According to Assumption 2, we can use decoy scores to estimate the distribution of the scores of true null target variables. Here we use the kernel density estimation for implementation:
	\begin{equation}\label{equ:f0t-est}
		\widehat{f_{0 t}}\left(S | h_{0}\right)=\widehat{f_{d}}\left(S | h_{0}\right)=\frac{1}{|\mathcal{D}| h_{0}} \sum_{j \in \mathcal{D}} K\left(\frac{S-S_{j}}{h_{0}}\right)
	\end{equation}
	where 
	the function $K(\cdot)$ represents the kernel function, which we choose to be the Gaussian kernel. $h_0$ is the bandwidth of the kernel density estimation and we use the decoy scores to select an  optimal bandwidth through cross validation.

	\subsection{$f_{1t}$ and $fdr_t$ estimation}
	$f_{1t}$ is the density of scores of non-null target variables. We use the target scores to estimate it following the iterative framework of kerfdr.
	First, for simplicity, we define a quantity as
	$$	p_{j}=\frac{\pi_{1 t} f_{1 t}\left(S_{j}\right)}{\pi_{0 t} f_{0 t}\left(S_{j }\right)+\pi_{1 t} f_{1 t}\left(S_{j}\right)}, \quad j \in \mathcal{T}  $$
	Obviously, the local false discovery rate of variable $j$ is
	$$	f d r_{t}(S_j)=1-p_{j}, \quad j \in \mathcal{T} $$
	
	In the iteration process, the aim is to estimate $f_{1t}$ and $p_j$'s 
	simultaneously. The process is described below:
	
	1. Initiation
	
	For the variable $j^*$ with the highest score, set $\widehat{p_{j^*}}^{(0)}$ to 1, and set $\widehat{p_{j}}^{(0)}, j \in \mathcal{T} \setminus \{j^*\}$ to 0. When there are ties in the highest scores, set all of the corresponding $\widehat{p_{j^*}}^{(0)}$'s to 1.
	
	2. Iteration
	
	(2.1) Estimation of $f_{1t}$
	
	We estimate $f_{1t}$ using the kernel density  estimation method as well. The estimate in the $l$-th iteration is
	$$	\widehat{f_{1 t}}^{(l)}\left(S | h_{1}\right)=\frac{\sum_{j \in  \mathcal{T}} \widehat{p_j}^{(l-1)} K\left(\frac{S-S_{j}}{h_{1}}\right)}{h_{1} \sum_{j \in \mathcal{T}} \widehat{p_j}^{(l-1)}}, \quad l \ge 1  $$
	where 
	the function $K(\cdot)$  is still chosen as Gaussian, and $h_1$ is optimized using the target scores by cross validation.
	
	(2.2) Updating $\widehat{p_j}$'s
	
	Having $\widehat{\pi_{0 t}}$, $\widehat{f_{0t}}$ and $\widehat{f_{1t}}$, we 
	update $\widehat{p_j}$ for variable $j$ as
	$$	\widehat{p_j}^{(l)}
		=\frac{\widehat{\pi_{1 t}} \widehat{f_{1 t}}^{(l)}(S_{j})}{\widehat{\pi_{0t}} \widehat{f_{0 t}}(S_{j})+\widehat{\pi_{1 t}} \widehat{f_{1 t}}^{(l)}(S_{j})}, \quad j \in \mathcal{T}, ~l \ge 1  $$
	
	3. Stopping criterion
	
	Stop the iteration if $l \ge l_{max}$ or
	$		\max _{j} \frac{\left|\widehat{p_j}^{(l)}-\widehat{p_j}^{(l-1)}\right|}{\widehat{p_j}^{(l-1)}}<\epsilon , ~ j \in \mathcal{T}$, where $l_{max}$ is the maximal number of iteration time and $\epsilon$ is a minor value which we choose as a threshold. Otherwise, go back to "Iteration" step.
	
	4. fdr estimation
	
	Finally, with the optimal $\widehat{p}_j$'s estimated, the objective fdr can be estimated as
		$$	\widehat{fdr}_{j}^{(l)}
			=1-\widehat{p_j}^{(l)} , \quad j \in \mathcal{T}$$

	The whole TDfdr algorithm is described in Algorithm 1. It takes the labels and scores of variables as the input and outputs the fdr estimates for target variables.

	\begin{algorithm}[H]
		\caption{TDfdr fdr estimation algorithm} 
		\hspace*{0.05in} {\bf Input:} 
		Labels $L_{j}$ and scores $S_{j}$ of variables, $j=1,2, \ldots , m$.\\
		\hspace*{0.05in} {\bf Output:} 
		fdr estimates for target variables
		\begin{algorithmic}[1]
			
			\State $\pi_{0t}$ estimation:\quad 
			$
			\widehat{\pi_{0 t}}=\frac{|\mathcal{D}|}{|\mathcal{T}|}={\frac{\#\left\{j=1,2, \ldots, m :~ L_{j}=D \right\}}{\#\left\{j=1,2, \ldots, m :~ L_{j}=T\right\}}}
			$, \quad
			$
			\widehat{\pi_{0 t}}=\min \left\{\widehat{\pi_{0 t}}, 1\right\}
			$
			
			\State $f_{0t}$ estimation:\quad
			$
			\widehat{f_{0 t}}\left(S | h_{0}\right)=\widehat{f_{d}}\left(S | h_{0}\right)=\frac{1}{|\mathcal{D}| h_{0}} \sum_{j \in \mathcal{D}} K\left(\frac{S-S_{j}}{h_{0}}\right)
			$
			
			\State Initiation:\quad
			$\widehat{p_{j^*}}^{(0)} = 1$, for $j^* = \mathop{\arg\max}\limits_{j \in \mathcal{T}}{S_j}$ and $\widehat{p_{j}}^{(0)}=0, ~j \in \mathcal{T} \setminus \{j^*\}.$
			~ $l=1$
			\Repeat
			\State estimate $f_{1t}$:\quad
			$\widehat{f_{1 t}}^{(l)}\left(S | h_{1}\right)=\frac{\sum_{j \in  \mathcal{T}} \widehat{p_j}^{(l-1)} K\left(\frac{S-S_{j}}{h_{1}}\right)}{h_{1} \sum_{j \in \mathcal{T}} \widehat{p_j}^{(l-1)}}, \quad l \ge 1$
			\State update $\widehat{p_j}$:\quad
			$\widehat{p_j}^{(l)}
			=\frac{\widehat{\pi_{1 t}} \widehat{f_{1 t}}^{(l)}(S_{j})}{\widehat{\pi_{0t}} \widehat{f_{0t}}(S_{j})+\widehat{\pi_{1 t}} \widehat{f_{1 t}}^{(l)}(S_{j})}, \quad j \in \mathcal{T}, ~l \ge 1$
			\State $l = l+1$
			\Until{$\max _{j} \frac{\left|\widehat{p_j}^{(l)}-\widehat{p_j}^{(l-1)}\right|}{\widehat{p_j}^{(l-1)}}<\epsilon , ~j \in \mathcal{T}$ or $l \ge l_{max}$}

			\State fdr estimation:\quad
			$\widehat{fdr}_{j}^{(l)} = 1-\widehat{p_j}^{(l)}, \quad j \in \mathcal{T}$ 
			
			\State \Return $\widehat{fdr}_{j}^{(l)}, \quad j \in \mathcal{T}$

		\end{algorithmic}
	\end{algorithm}

	\section{Simulation}\label{sec:simulation}
	\setcounter{equation}{0}
	
	To demonstrate the effect of our TDfdr method, we carried out simulations on the two-group study and the regression model. As for the fdr estimation methods, we mainly compared TDfdr with the locfdr method \cite{Efron:2002}.
	Moreover, TDFDR \cite{He:2018new} and the  knockoff filter \cite{knockoff:2015} were also tested when performing comparisons in terms of FDR.
	
	A complete simulation contains (1) generating random samples according to the predefined parameters; (2) computing labels or scores from the generated samples using a competition procedure;
	and (3) estimating the fdr of variables based on the computed statistics. 
	We repeated the simulation for $M$ times and compared the average results, including the estimation accuracies of $\pi_0$ and fdr, FDR control performance, and power. Detailed simulation designs and parameters of methods are described in Supplementary S2.

	\subsection{Performance evaluation}
	
	For TDfdr, we used the  $t$-statistics as input and estimated the fdr of target variables. 
	For locfdr, we first tested the standard locfdr method by inputting the $t$-statistics directly, and estimated the fdr of all variables, which we call the \itshape locfdr- \upshape method.  
	Then, 
	a transformation 
	was made to the statistics
	to better satisfy the assumption of "normal distribution under null hypothesis".
	The locfdr with this kind of transformation is referred to as the \itshape locfdr+ \upshape method. 
	In summary, the methods for fdr estimation were TDfdr, locfdr-, and locfdr+.

	In the simulations, we first compared $\pi_0$ estimations by TDfdr and locfdr. TDfdr estimates the $\pi_0$ as in Equation \ref{equ:pi0}.
	For locfdr we use the submethod of maximum likelihood (nulltype=1, the default) in the
	algorithm to obtain the $\pi_0$ estimation.
	
	Regarding the accuracy of fdr estimation, we used the averaged sample RMSE (root mean squared error) of $M$ repetitive simulations as the metric for comparison.
	The estimation RMSE for the $k$-th repetition is
	\begin{equation}\label{equ:fdrRMSE}
		RMSE^{(k)}=\sqrt{\frac{1}{|\mathcal{I}^{(k)}|} \sum_{j \in \mathcal{I}^{(k)}} \left(\widehat{fdr}_{j}^{(k)}- fd r_{j}^{(k)}\right)^{2}}, 
	\end{equation}
	where $\widehat{fdr}_{j}^{(k)}$ and $fdr_j^{(k)}$ are the estimated and real fdr  for the variable $j$ in repetition $k$, respectively. 
	$\mathcal{I}^{(k)}$ represents the variable set of interest (all variables or the target variables) in the $k$-th repetition.
	Then, the average $RMSE$ over all repetitions, i.e.,
		$$
		R M S E_{ave}=\frac{1}{M} \sum_{k=1}^{M}R M S E^{(k)},
	$$
	was used as the performance metric.
	
	As mentioned before, TDfdr 
	focuses only on the fdr of target variables, i.e., $fdr_t$. Therefore, we need to know the real values of $fdr_t$ as given in Equation \ref{equ:realfdrt}, for comparison with the estimated fdr.
	With all the data simulated artificially, whether a variable belongs to the null set is known, and thus the real value of $\pi_{0t}$ can be directly computed according to Equation \ref{equ:realpi0t}.
	Next the real $f_{0t}$ and $f_{1t}$
	are computed through kernel density estimation using the 
	final scores of target variables with known labels.
	Specifically, we use the final scores of the null target variables to estimate the real target null density $f_{0t}$ and the final scores of non-null target  variables to estimate the real target non-null density $f_{1t}$.
	That is, 
	$$	f_{0 t}\left(S | h_{0t}\right)=\frac{1}{|\mathcal{T}_0| h_{0t}} \sum_{j \in \mathcal{T}_0} K\left(\frac{S-S_j}{h_{0t}}\right) $$
	$$	f_{1t}\left(S | h_{1t}\right)=\frac{1}{|\mathcal{T}_1| h_{1t}} \sum_{j \in \mathcal{T}_1} K\left(\frac{S-S_j}{h_{1t}}\right) $$
	where the explanations of parameters are analogous to those in Equation \ref{equ:f0t-est}.
	The values of fdr are computed similarly with regard to all variables instead of target ones.
	
	In addition, we also evaluated our method in terms of FDR.
	Based on the connection between fdr and FDR in Equation \ref{fdr2FDR}, 
	we can estimate FDR practically from the estimated fdr as 
		$	\widehat{FDR}(S) = mean\{\widehat{fdr_j}:S_j < S\} , ~j \in \mathcal{I}, $ 
	where the meaning of $\mathcal{I}$ is the same as that in Equation \ref{equ:fdrRMSE}.

	\subsection{Simulation results on two-group study}
	
	The simulation results are arranged in three parts, i.e., $\pi_0$ estimation, fdr estimation, and FDR related results. For the latter two, we only show the results when $\pi_0=0.8$ in the main text, and more results are given in Supplementary S3.1.

\subsubsection{Results of $\pi_{0}$ estimation}

Due to the importance 
of $\pi_0$ in fdr estimation, we first compared the accuracy of $\pi_0$ estimation by the TDfdr and locfdr methods.

\begin{figure}[H]
	\centering
	\subfigure[Estimated vs. real values of $\pi_0$ of normal data]{
		\begin{minipage}[t]{\linewidth}
			\centering
			\includegraphics[width=0.9\textwidth]{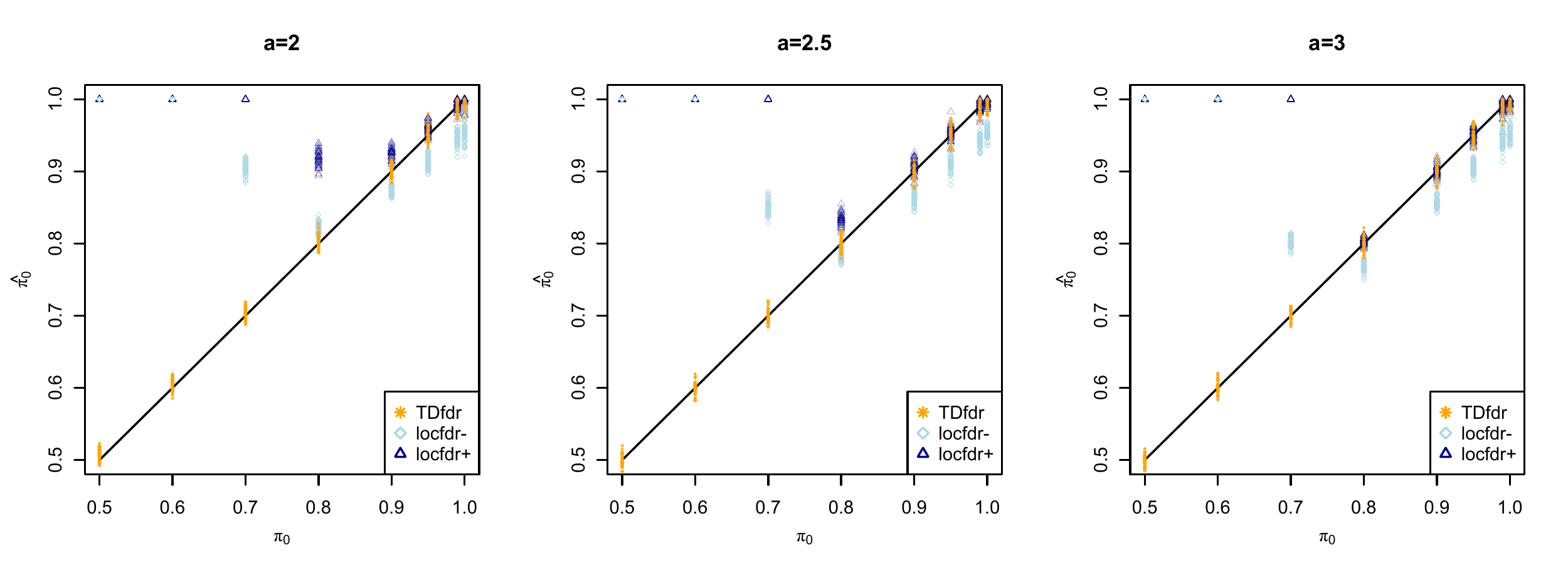}
			\label{subfig:pi01}
		\end{minipage}%
	}%
	\\
	\subfigure[Estimated vs. real values of $\pi_0$ of gamma data]{
		\begin{minipage}[t]{\linewidth}
			\centering
			\includegraphics[width=0.9\textwidth]{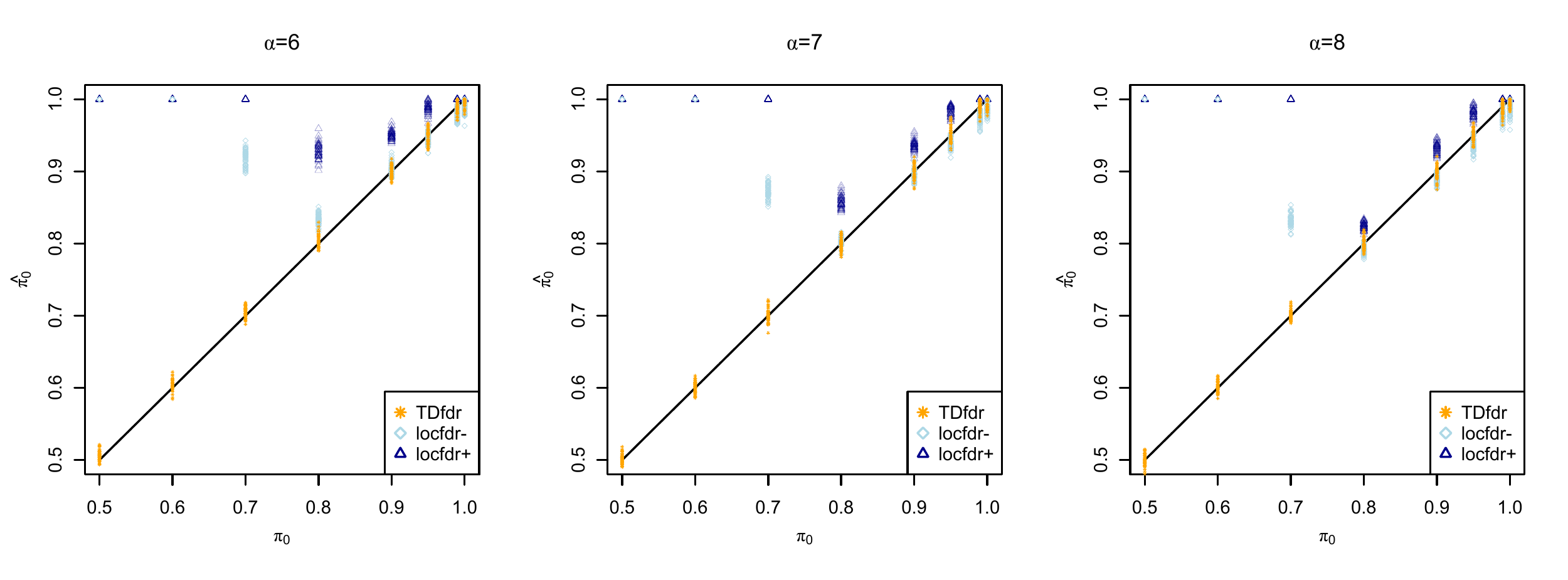}
			\label{subfig:pi02}
		\end{minipage}
	}
	\\
	\subfigure[RMSEs of fdr estimation of normal data ($\pi_0=0.8$)]{
		\begin{minipage}[t]{\linewidth}
			\centering
			\includegraphics[width=0.9\textwidth,height=4.5cm]{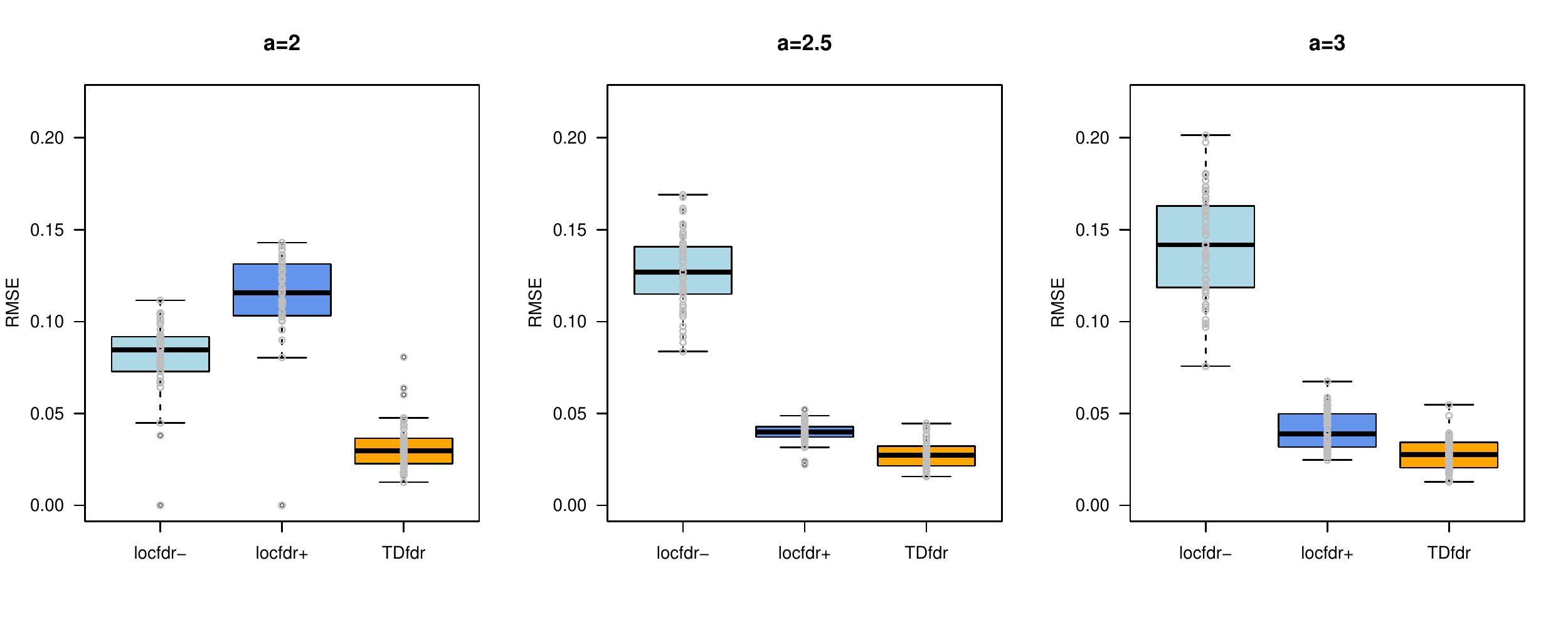}
			\label{subfig:fdr1}
		\end{minipage}%
	}
	\\
	\subfigure[RMSEs of fdr estimation of gamma data ($\pi_0=0.8$)]{
		\begin{minipage}[t]{\linewidth}
			\centering
			\includegraphics[width=0.9\textwidth,height=4.5cm]{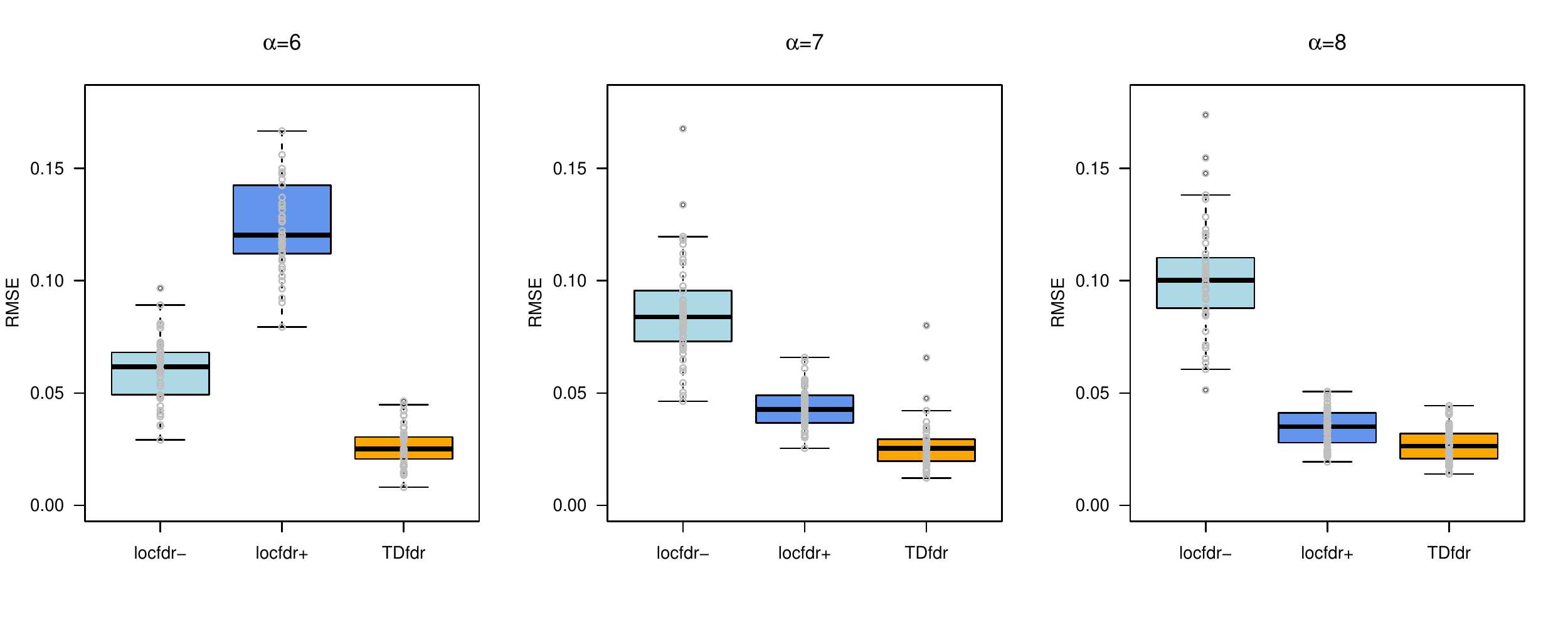}
			\label{subfig:fdr2}
		\end{minipage}%
	}
	\caption{$\pi_{0}$ and fdr estimation results of two-group data}
	\setlength{\belowcaptionskip}{-0.cm}
	\label{fig:pi0andfdr}
\end{figure}

Note that $\pi_{0t}$ could not be compared because locfdr works on all variables instead of target ones.
We chose a series of real $\pi_0$ values for simulation, and obtained $\pi_0$ estimates using the TDfdr and locfdr methods.
Two $\pi_0$ estimates were obtained by the locfdr method, 
locfdr- and locfdr+, respectively. 
For the real values of $\pi_0$, we chose $\pi_0$ = 0.5, 0.6, 0.7, 0.8, 0.9, 0.95, 0.99 and 1. The scatter plots of $\pi_0$ estimates for normal and gamma data are shown in Figure \ref{fig:pi0andfdr}(a) and (b).


It is shown 
that TDfdr estimated $\pi_0$ more accurately 
than locfdr, especially in the cases where the real $\pi_0$ was less than 0.8. 
In most cases,  
TDfdr achieved the highest 
accuracy. 
Moreover, TDfdr kept a decent level of variances of estimations.

\subsubsection{Results of fdr estimation}

The boxplots of fdr estimation RMSEs of the normal and gamma data ($\pi_0=0.8$) are shown in Figure \ref{fig:pi0andfdr}(c) and (d)
, respectively.
For the results corresponding to other values of $\pi_0$, see Supplementary Figures \ref{fig:normalSimu-fdr} and \ref{fig:gammaSimu-fdr}.


As Figure
\ref{fig:pi0andfdr}(c)
shows, 
locfdr+ has smaller RMSEs than locfdr-, indicating that 
the transformation of input scores increased the accuracy of locfdr.
In the comparison between TDfdr and locfdr, the RMSEs of TDfdr are significantly smaller than those of both locfdr- and locfdr+. 
It can also be observed from Supplementary Figure \ref{fig:normalSimu-fdr} that the RMSEs of TDfdr decrease as the group difference 
($a$) 
increases, while the RMSEs of locfdr estimation exhibit no obvious trend towards the group difference.
Overall, the fdr estimation by TDfdr is more accurate and stable than locfdr for normal data.

For gamma data, the boxplots of fdr RMSEs in Figure \ref{fig:pi0andfdr}(d) show a similar trend to normal data. 
The RMSEs of TDfdr are much lower than those of locfdr-, and are lower than or comparable to those of locfdr+.

As shown in Supplementary Figures \ref{fig:normalSimu-fdr} and \ref{fig:gammaSimu-fdr}, 
locfdr+ is not as stable as TDfdr, which becomes more clear 
in cases where the null proportion is small (0.8), 
for both normal and gamma data. 
In contrast, the medians of RMSEs of TDfdr remain less than 0.05 in all cases, demonstrating its robustness and stability to various conditions.

\subsubsection{Results of FDR comparison}
With the FDR calculated from fdr, 
the effect of TDfdr were also evaluated in terms of FDR control and power.
For FDR control, we calculated the realized FDR of rejected variables as the mean of observed false discovery proportions (FDPs) in all repetitions, and then drew plots of realized FDR 
vs. FDR control threshold, to see whether different  methods are able to control FDR under varying thresholds. 
Points lying under the dashed line $x=y$ represent good control of FDR.


\begin{figure}[H]
\centering
\subfigure[FDR control results of normal data]{
	\begin{minipage}[t]{\linewidth}
		\centering
		\includegraphics[width=\textwidth]{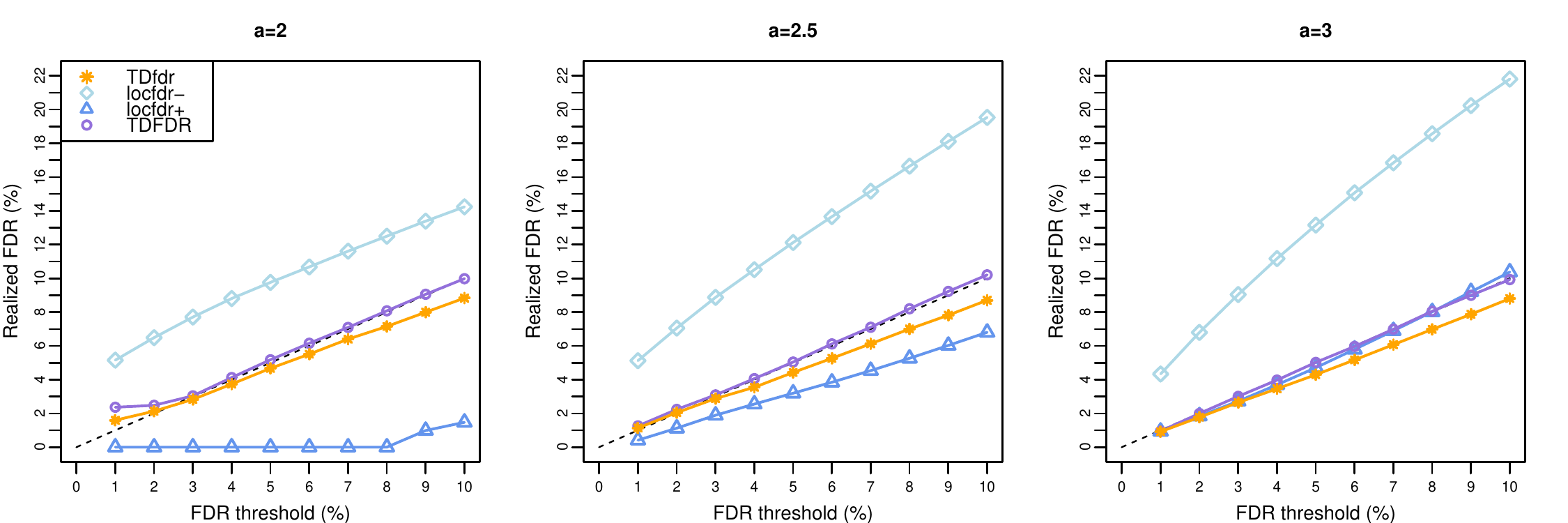}
		\label{subfig:control1}
	\end{minipage}%
}
\\
\subfigure[FDR control results of gamma data]{
	\begin{minipage}[t]{\linewidth}
		\centering
		\includegraphics[width=\textwidth]{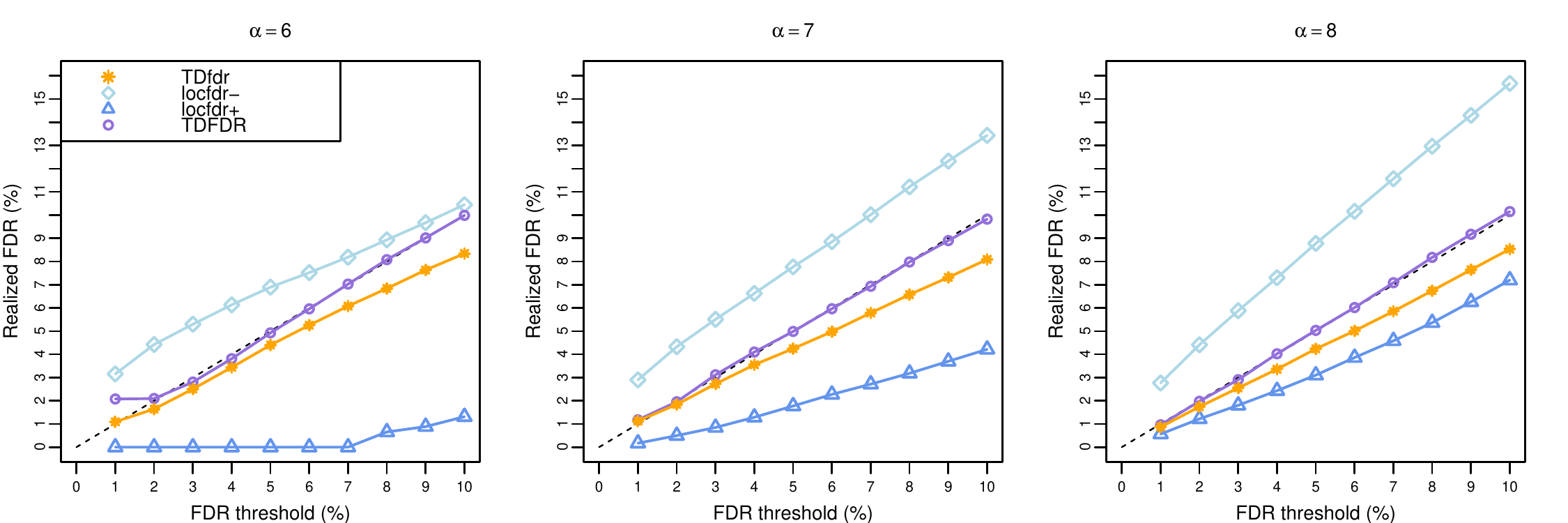}
		\label{subfig:control2}
	\end{minipage}%
}
\\
\subfigure[Powers of normal data]{
	\begin{minipage}[t]{\linewidth}
		\centering
		\includegraphics[width=\textwidth]{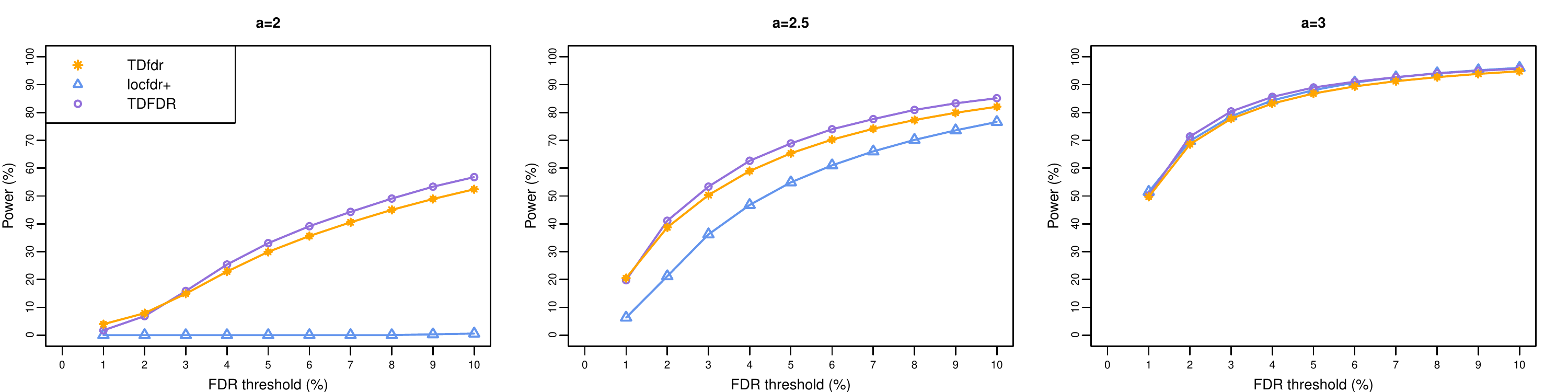}
		\label{subfig:power1}
	\end{minipage}%
}
\\
\subfigure[Powers of gamma data]{
	\begin{minipage}[t]{\linewidth}
		\centering
		\includegraphics[width=\textwidth]{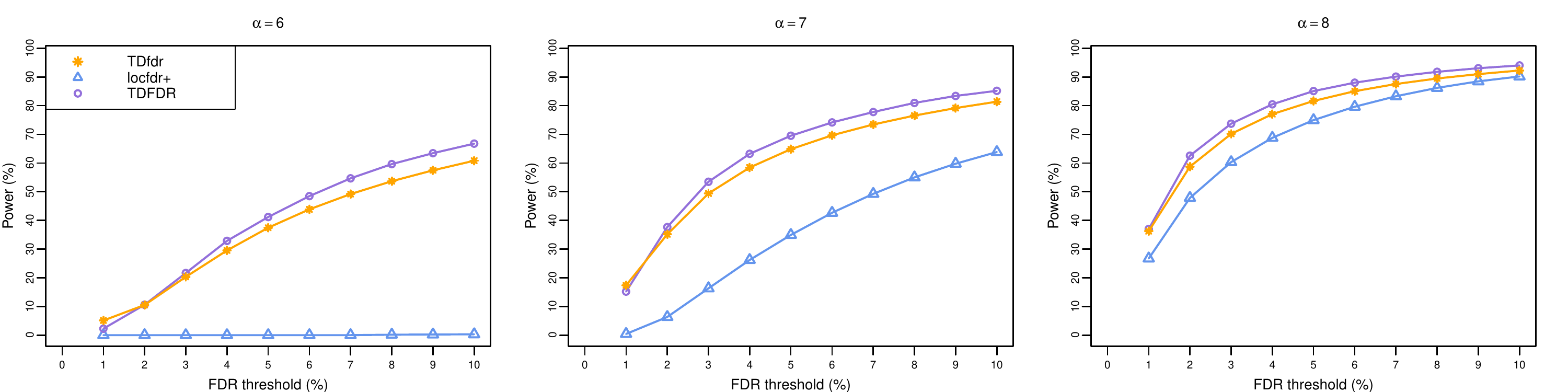}
		\label{subfig:power2}
	\end{minipage}%
}
\caption{FDR control results and powers of two-group data  ($\pi_0=0.8$)}
\label{fig:controlandpower}
\end{figure}

For normal data, Figure  
\ref{fig:controlandpower}(a) and Supplementary Figure \ref{fig:normalSimu-control}
show that locfdr-
seriously failed to control FDR, 
while the remaining three methods (TDfdr, locfdr+ and TDFDR) yielded acceptable control of FDR. 
More specifically, locfdr+ performed conservatively in some cases, and in contrast, both TDfdr and TDFDR obtained realized FDRs closer to the nominal ones. 
For the cases where the group difference $a=2$ and the FDR thresholds were small, these three methods all yielded poor controlling results.

For gamma data, 
similar results are observed 
in Figure \ref{fig:controlandpower}(b) and Supplementary Figures 
\ref{fig:gammaSimu-control}. 
Both TDfdr and locfdr+ achieved decent FDR control, while TDfdr realized the FDR closer to the thresholds than locfdr+. TDFDR behaved even more liberal than TDfdr, controlling the FDR to the exact level of nominal ones. 


In general, among all the methods, TDfdr and TDFDR are the two which controlled the FDR most closely to the given thresholds, though in some cases the FDR slightly got out of control. However, the realized FDR by locfdr- deviated far from the thresholds, meaning serious failure in FDR control. For locfdr+, it obviously estimated the FDR too conservatively, in spite of its best performance in FDR control.

The average powers for normal data are plotted in 
Figure \ref{fig:controlandpower}(c) and Supplementary Figure \ref{fig:normalSimu-power}.
Due to the uncontrollable FDR by locfdr-, we here only display the powers of the remaining three methods.
TDfdr performed decently in most of the situations, with robustness to different configurations.
Compared to locfdr+, TDfdr obtained comparable or better results in most cases, while became slightly worse in some easier cases where the group difference was larger (shown in the bottom right of Supplementary Figure \ref{fig:normalSimu-power}). It is worth noting that locfdr+ output zero power in the case where $a=2$ \& $\pi_0 = 0.8$, which means a failure in selecting significant variables for all the thresholds.

For gamma data, 
Figure \ref{fig:controlandpower}(d) and Supplementary Figure \ref{fig:gammaSimu-power}
show that TDfdr produced powers all surpassing locfdr+. Meanwhile, locfdr+ failed again in the hardest case (when $\alpha =6$ \& $\pi_0=0.8$).

To sum up, 
TDfdr showed higher power, especially for gamma data. Locfdr+ achieved slightly superior performances to TDfdr in a few cases, but from the overall perspective, it was less stable to produce valid variable selections.

\subsection{Simulation results on knockoff-based variable selection}

As a competition-based procedure, knockoff filter calculates statistics with different signs and absolute values, which play the similar roles of labels and final scores, respectively, in the framework of TDFDR. 
We used the statistics of knockoff filter as the input of TDfdr and locfdr to estimate fdr, and demonstrated the universality of our method on competition-based procedures. 
In addition, we evaluated the FDR results in comparison with knockoff filter, illustrating the ability of our method in FDR control and variable selection. 
Again, we show here only the results of $\pi_0=0.8$ for fdr estimation and FDR control \& power, and give more results in Supplementary S3.2.
	
	\subsubsection{Results of $\pi_{0t}$ estimation}
\begin{figure}[H]
		\centering
		\subfigure[Estimated vs. real values of $\pi_{0t}$ of regression data (independent cases)]{
			\begin{minipage}[t]{\linewidth}
				\centering
				\includegraphics[width=0.9\textwidth]{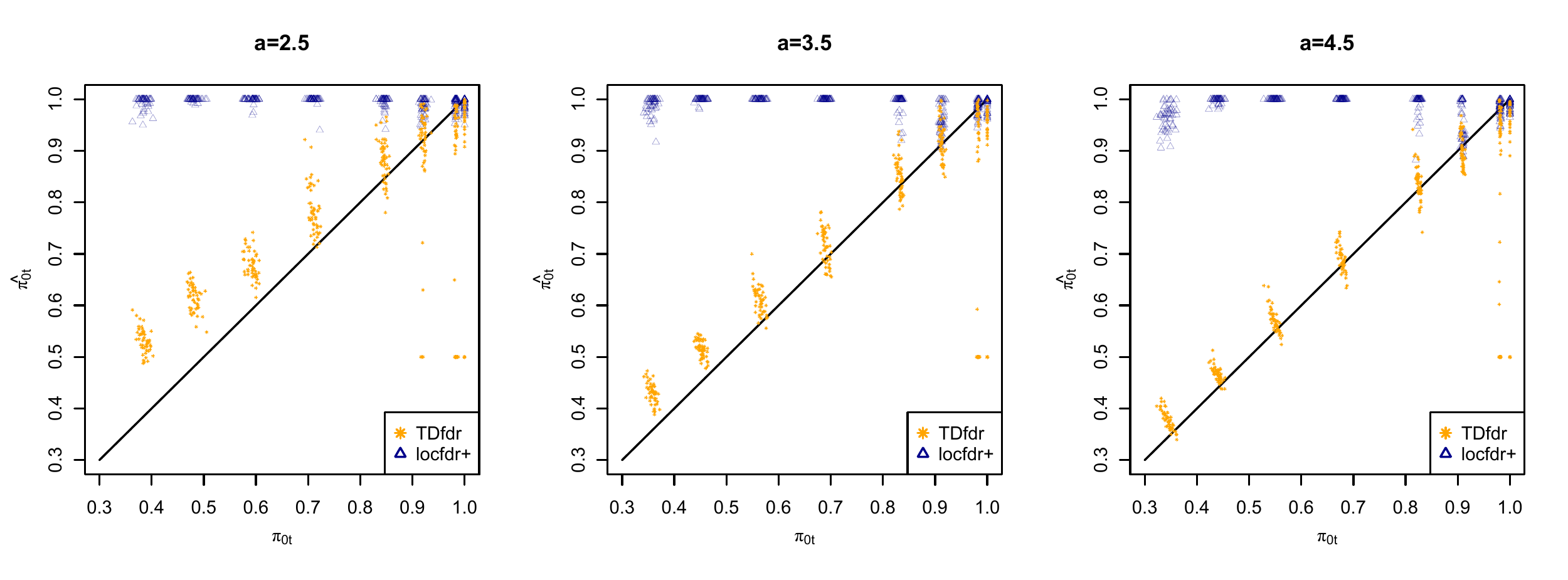}
				\label{subfig:KOsimu-pi0t1}
			\end{minipage}
		}
		\\
		\subfigure[Estimated vs. real values of $\pi_{0t}$ of regression data (dependent cases)]{
			\begin{minipage}[t]{\linewidth}
				\centering
				\includegraphics[width=0.9\textwidth]{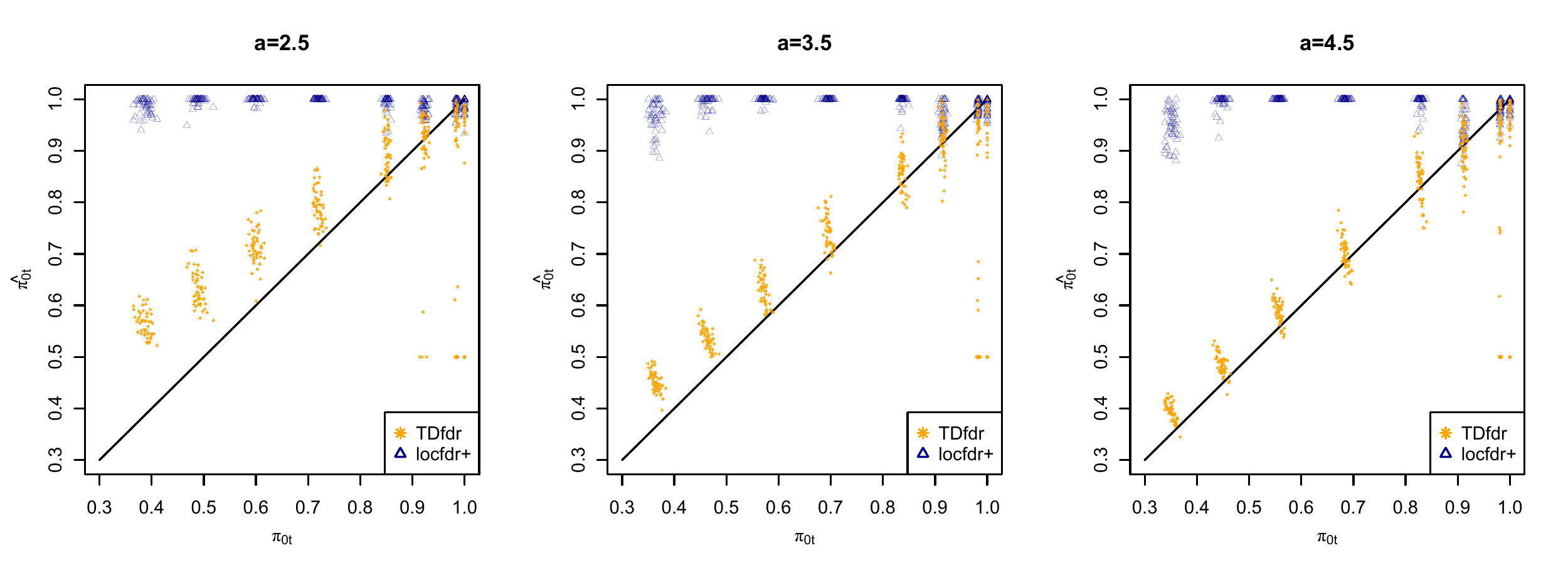}
				\label{subfig:KOsimu-pi0t2}
			\end{minipage}
		}
		\\
		\subfigure[RMSEs of fdr estimation of regression data (independent cases, $\pi_0$=0.8)]{
			\begin{minipage}[t]{1\linewidth}
				\centering
				\includegraphics[width=0.9\textwidth, height=4.5cm]{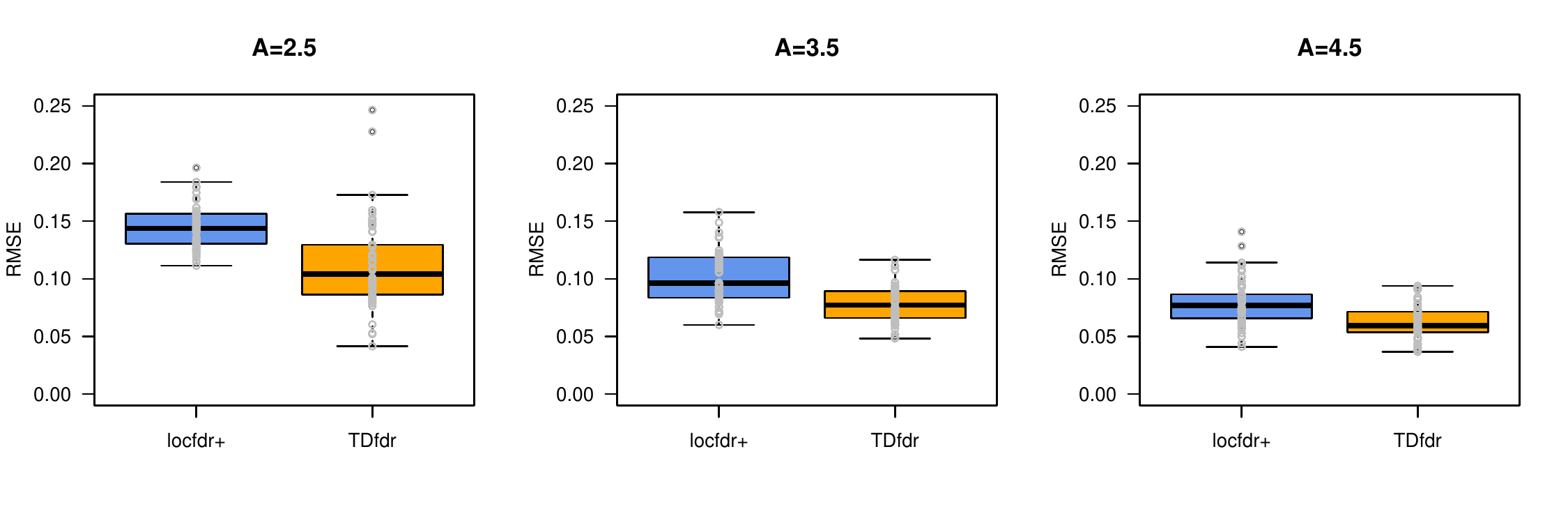}
				\label{subfig:KOfdr1}
			\end{minipage}
		}
		\\
		\subfigure[RMSEs of fdr estimation of regression data (dependent cases, $\pi_0$=0.8)]{
			\begin{minipage}[t]{1\linewidth}
				\centering
				\includegraphics[width=0.9\textwidth, height=4.5cm]{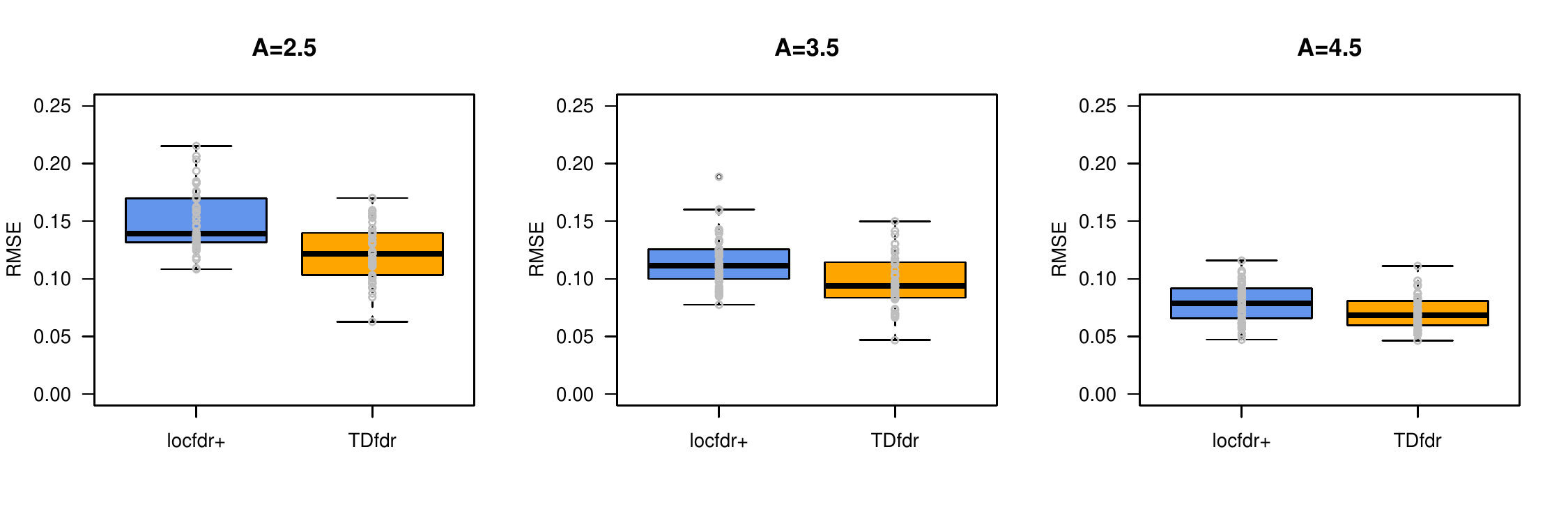}
				\label{subfig:KOfdr2}
			\end{minipage}
		}
		\caption{$\pi_{0t}$ and fdr estimation results of regression data}
		\label{fig:KOsimu-pi0tandfdr}
	\end{figure}

	Figure
	\ref{fig:KOsimu-pi0tandfdr}%
	(a) 
	and (b)
	corresponding to independent and dependent cases respectively,
	shows the $\pi_{0t}$ estimation results of TDfdr and locfdr+. It can be seen that locfdr+ mistakenly estimated $\pi_{0t}$ to be one, while the estimation by TDfdr was much more accurate. 
	
	
	\subsubsection{Results of fdr comparison}
	
	Figure \ref{fig:KOsimu-pi0tandfdr}(c) and Supplementary Figure \ref{fig:koSimu-fdr} compare the fdr estimation RMSEs of TDfdr and locfdr+ for independent variables. For all the three signal amplitudes ($A$) when $\pi_0=0.8$, TDfdr yielded more accurate estimations than locfdr+. 
	The full results in Supplementary Figure \ref{fig:koSimu-fdr} show that for all the cases but one ($A=2.5$ \& $\pi_0=0.9$), TDfdr estimated fdr with less median error, demonstrating superior performance to locfdr+.

	Figure \ref{fig:KOsimu-pi0tandfdr}(d) and Supplementary Figure \ref{fig:koSimu-fdr-corr} show the comparison of fdr for variables with dependency. 
	In the existence of dependency, 
	the number of wins of TDfdr against locfdr+ decreased compared to the independent cases. However, the advantage of TDfdr over locfdr+ is still obvious overall, in spite of the slight inferiority in some cases to locfdr+. 
	
	In brief, these results demonstrate that for most configurations of null proportions and signal amplitudes, TDfdr yielded more accurate fdr estimation than locfdr.

	\subsubsection{Results of FDR comparison}
	
	For FDR, we compared the deduced FDRs from fdr estimated by TDfdr and locfdr, with the FDR given by knockoff filter. 
	
	\begin{figure}[H]
		\setcounter{subfigure}{0}
		\centering
		\subfigure[FDR control results of regression data (independent cases)]{
			\begin{minipage}[t]{\linewidth}
				\centering
				\includegraphics[width=0.9\textwidth]{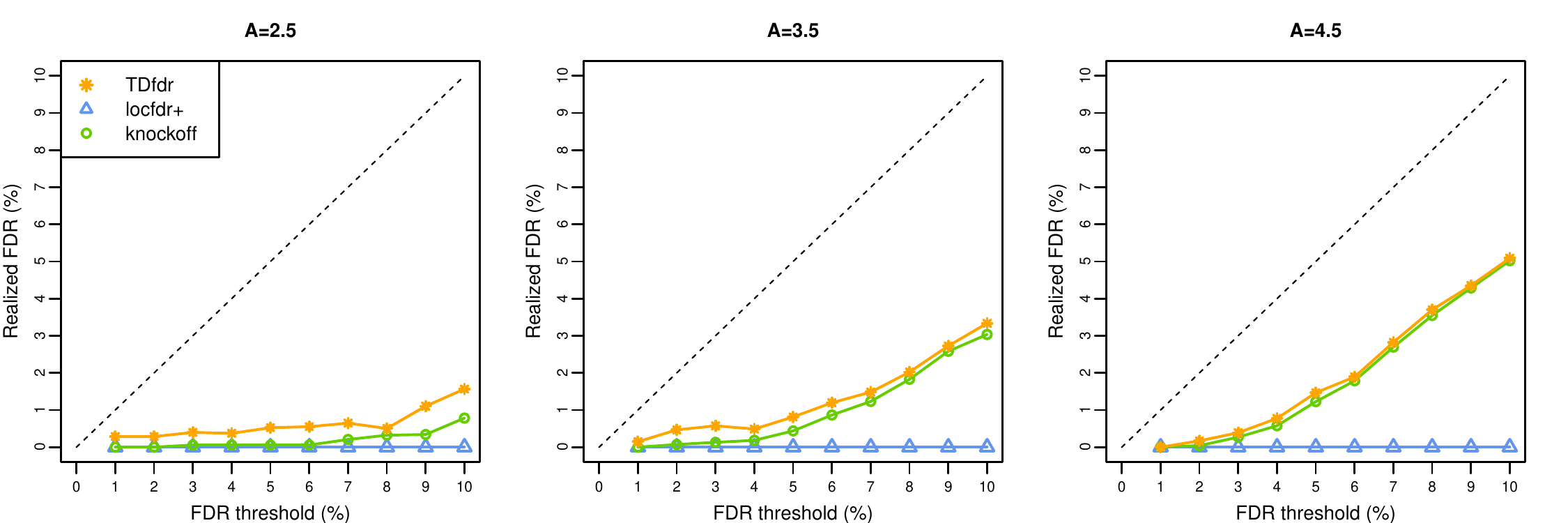}
				\label{subfig:KOcontrol1}
			\end{minipage}
		}
		\\
		\subfigure[FDR control results of regression data (dependent cases)]{
			\begin{minipage}[t]{\linewidth}
				\centering
				\includegraphics[width=0.9\textwidth]{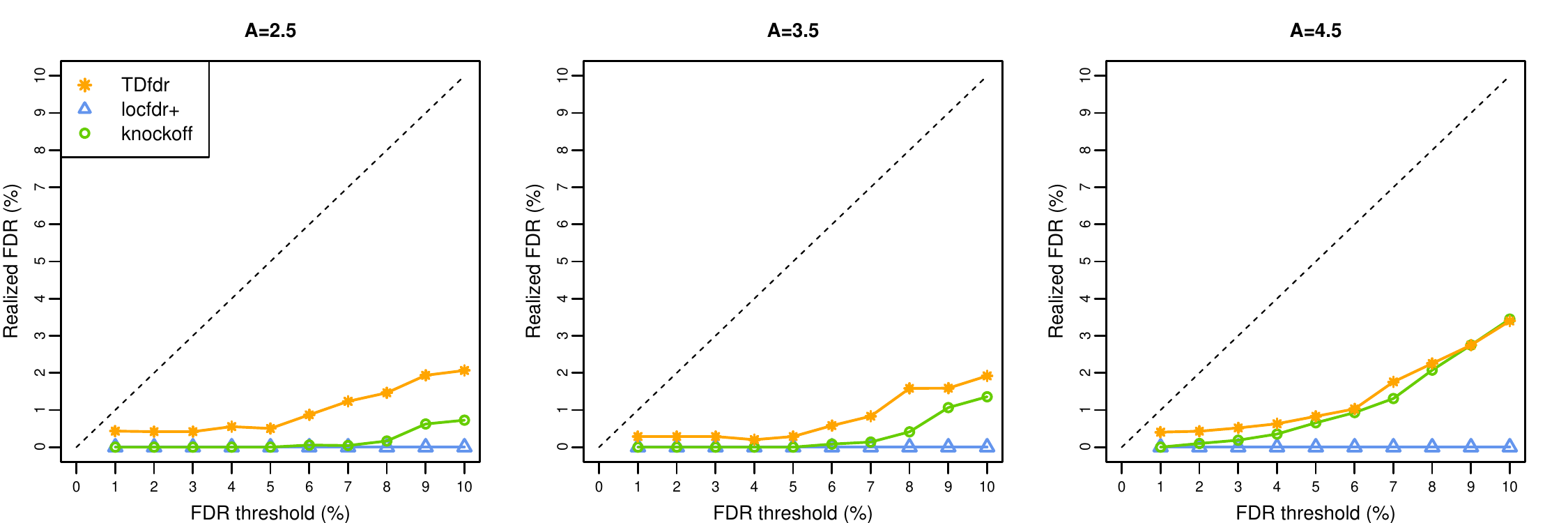}
				\label{subfig:KOcontrol2}
			\end{minipage}
		}
		\\
		\subfigure[Powers of regression data (independent cases)]{
			\begin{minipage}[t]{\linewidth}
				\centering
				\includegraphics[width=0.9\textwidth]{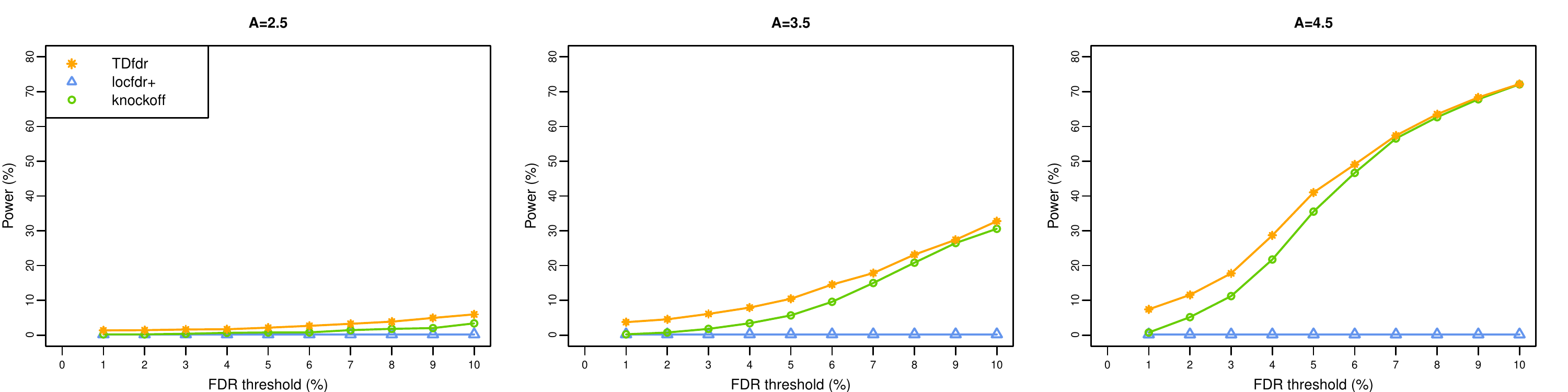}
				\label{subfig:KOpower1}
			\end{minipage}%
		}%
		\\
		\subfigure[Powers of regression data (dependent cases)]{
			\begin{minipage}[t]{\linewidth}
				\centering
				\includegraphics[width=0.9\textwidth]{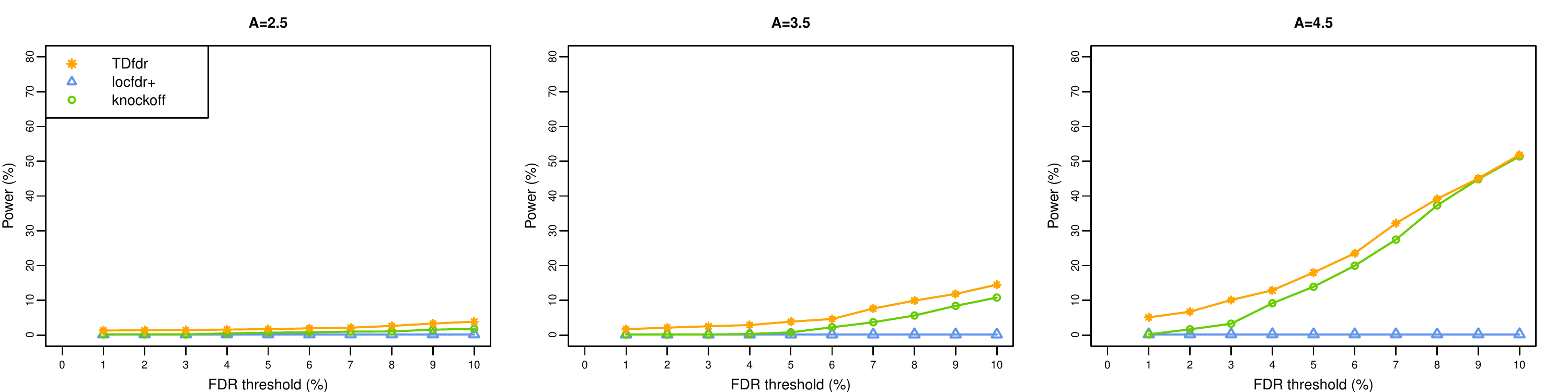}
				\label{subfig:KOpower2}
			\end{minipage}%
		}%
		\centering
		\caption{FDR control results and powers of regression data ($\pi_0=0.8$)}
		\label{fig:KOcontrolandpower}
	\end{figure}
	
	As shown in Figure \ref{fig:KOcontrolandpower}(a)
	and 
	(b), for different FDR thresholds, the three methods all succeeded in controlling FDR when $\pi_0=0.8$, even with the dependency between variables. Among the three methods, TDfdr controlled FDR most liberally, with the lines of realized FDR closest to the line $x=y$. In contrast, knockoff filter and locfdr+, especially the latter, were too conservative in FDR control. This trend directly led to a higher power of TDfdr.


	Regarding power, for all the thresholds in both independent (Figure \ref{fig:KOcontrolandpower}(c)) and dependent cases (Figure \ref{fig:KOcontrolandpower}(d)), TDfdr achieved the highest power. 
	Locfdr+ was barely able to select out any significant variables, as shown by the power line around zero. This also explained why its results of control FDR were so close to zero. 
	Note that the FDR is by definition zero when the selection set is empty. 
	
	When $\pi_0=0.9$ and $0.95$ (Supplementary Figures \ref{fig:koSimu-control}-\ref{fig:koSimu-power-corr}), we observed similar results. TDfdr still had higher powers than locfdr+ and knockoff filter, and again, locfdr+ selected few significant variables. TDfdr controlled FDR well in all cases except when $\pi_0=0.95$ and FDR threshold was 1-2\% (independent cases).
	
	\section{Real data analysis}\label{sec:realdata}
	
	We applied the TDfdr method to two real  datasets, including a two-sample COVID-19 dataset and a regression dataset of HIV drug resistance, and compared it with other methods.
	
	\subsection{COVID-19 data}
	
	We utilized a dataset from samples of COVID-19 sera to evaluate the performance of TDfdr. The results from the original paper \cite{Guo:2020} were employed as a reference. Moreover, the locfdr method (including locfdr- and locfdr+) was tested for comparison.
	
	The dataset was from a study of proteomic 
	characterization of COVID-19 patient sera \cite{Guo:2020}, and contains the measurements of 894 proteins 
	in the serum samples of 93 subjects divided into four groups. 
	Details of sample grouping are listed in Supplementary Table \ref{table:samplegroup}. 
	Missing data were processed by first deleting the all-missing proteins, and then filling the remaining missing values with zeros. As a result, the number of variables (proteins) was trimmed from 894 to 791.
	
	To identify the significantly differentially expressed proteins related to the COVID-19 disease, the original paper firstly employed three case-control comparisons with the "Healthy" group serving as the control, i.e., (1) Severe vs. Healthy, (2) Nonsevere vs. Healthy, and (3) Non-COVID-19 vs. Healthy, and then reported the final proteins as the union of the first two comparisons excluding the third. For the original results, 105 differentially expressed proteins for COVID-19 patients were discovered 
	using the combined criteria of FDR and fold-change. 
	
	Referring to the original paper, we used the 5\% FDR threshold and replaced the FDR estimation method with TDfdr and locfdr deduced FDR.
	Due to the randomness inside TDfdr, we repeated its procedure for 49 times, and calculated the median number of selected proteins. Both locfdr- and locfdr+ were run once, as they provide deterministic results. The number of proteins selected through the three aforementioned comparisons, the number of reported proteins, and the number of their intersections with the original 105 proteins were given in Table \ref{table:COVID19number} for comparison.
	\begin{table}[H]
		\scriptsize
		\setlength{\abovecaptionskip}{0pt}
		\setlength{\belowcaptionskip}{0pt}
		\captionsetup{font={small}}
		\caption{Numbers of selected proteins by different methods and their intersections with the original result}
		\label{table:COVID19number}
		\begin{center}
			\scalebox{1}{
				\begin{threeparttable}
					\begin{tabular}{p{35pt}p{45pt}p{45pt}p{60pt}p{30pt}p{90pt}}
						\hline
						Methods & Severe ~~~~~~ vs. Healthy & Nonsevere vs. Healthy & Non-COVID-19 vs. Healthy & Final report$^{*}$  & Intersection of final report with original proteins\\
						\hline

						Original & 120 & 43 & 28 & 105 & -\\
						TDfdr & 143 & 52 & 31 & 124  & 104 \\
						locfdr- & 153 & 83 & 26 &
						144 & 104 \\
						locfdr+ & 130 & 71 & 3 &
						144 & 103 \\
						\hline
					\end{tabular}
					\begin{tablenotes}
						\scriptsize
						\item[*] Final reported proteins = (Severe vs. Healthy) $\cup$ (Nonsevere vs. Healthy) $\backslash$ (Non-COVID-19 vs. Healthy)
					\end{tablenotes}
			\end{threeparttable}}
		\end{center}
	\end{table}
	
	In the individual case-control comparisons, TDfdr and locfdr methods generally selected more significant proteins than the original result, except that locfdr+ reported only 3 proteins in the "Non-COVID-19 vs. Healthy" comparison. Regarding the final reported proteins,
	locfdr- and locfdr+ both selected 144 proteins, and TDfdr selected 124. However, after intersecting with the original results, TDfdr and locfdr- both obtained 104 overlapping proteins, very close to the all 105. 
	This implies that TDfdr may possess a lower real FDR than locfdr.

	Further,
	we analysed the intersection and the difference of the protein sets detected through TDfdr and the original method. 
	TDfdr was run 3 times, and the intersecting proteins were believed to be high confidence and were subjected to analysis. 
	We compared four sets of proteins through the Venn plot which is shown in Supplementary Figure \ref{fig:Venn}.  
	For the proteins specific to the severe COVID-19 patients, the original paper reported 97 significant ones, while TDfdr reported 123 with high confidence, as shown in the areas of "Orig(S-)" and "TDfdr(S-)", respectively, in the Venn plot. In the 97 proteins, 96 were included in the results of TDfdr.
	Moreover, TDfdr found another 20 severe proteins which were not reported in the original study. 
	For the differentially expressed proteins in the nonsevere COVID-19 group (marked as "TDfdr(N-)" and "Orig(N-)"), TDfdr's result covered all the 33 proteins that were found in the original paper, with 6 newly reported ones.
	
	Regarding the 123 significant severe-COVID-19 proteins selected by TDfdr, we carried out pathway analysis to demonstrate their biological functions in the pathways. 
	The R package "clusterProfiler" was utilized to search for important pathways and make visualizations.  Supplementary Figure \ref{fig:pathway}
	shows the results, where the x-axis of the plot represents the number of proteins in the corresponding pathways.
	The 20 most significant pathways are displayed,
	and the small values of the adjusted $p$-values demonstrate that these pathways were enriched from the genes with high confidence. 
	Among these significant pathways, two were found concordant with those reported in the original paper \cite{Guo:2020}. Specifically, the pathway "GOBP platelet degranulation" in Supplementary Figure \ref{fig:pathway} 
	corresponds exactly to the "platelet degranulation" in the original paper, and the "GOBP complement activation" is similar to the "complement system" in the original paper.
	The third pathway enriched in the original paper is called "Macrophage function", which is closely related to immune response, and it also has a corresponding pathway in our results called "GOBP humoral immune responds". In summary, the three pathway clusters found in the original paper can all correspond to the ones in our analysis results.

	\subsection{HIV data}
	
	To demonstrate the effect of TDfdr on the  regression model, we employed it on an HIV dataset, which was also utilized to evaluate the knockoff method \cite{knockoff:2015}.
	
	As described in the original paper \cite{knockoff:2015}, 
	the dataset consists of drug resistance measurements and genotype information from samples of Human Immunodeficiency Virus
	Type 1 (HIV-1). The task is to detect mutations in the genotypes of HIV-1 that are associated with drug resistance.
	Specifically, the response variable $y_i$ is given by the log-fold increase of lab-tested drug resistance in the $i$-th sample, and the design matrix $X$ has entries $X_{ij} \in \{0, 1\}$, indicating presence or absence of mutation $j$ in the $i$-th sample.
	
\begin{figure}[H]
		\centering
		\subfigure[FDR threshold = 1\%]{
			\includegraphics[width=1\textwidth]{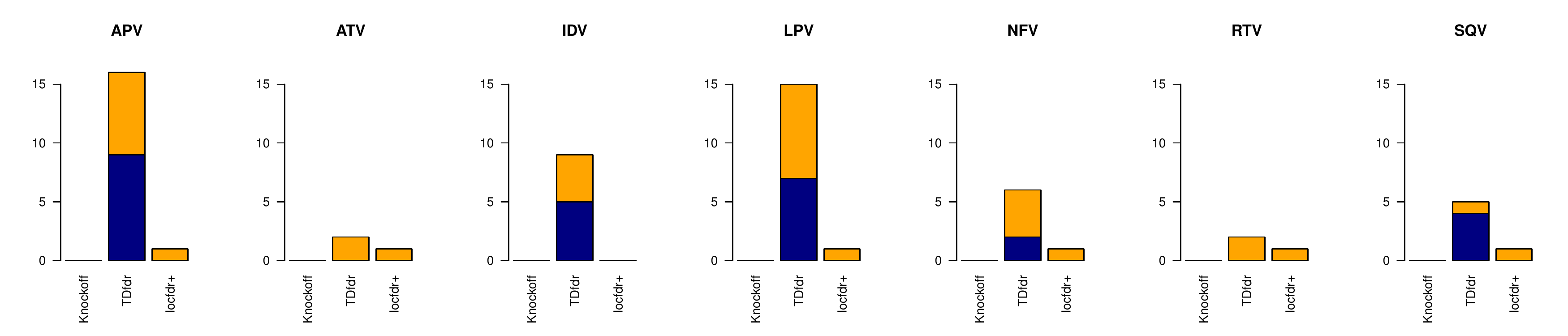}
		}
		\subfigure[FDR threshold = 5\%]{
			\includegraphics[width=1\textwidth]{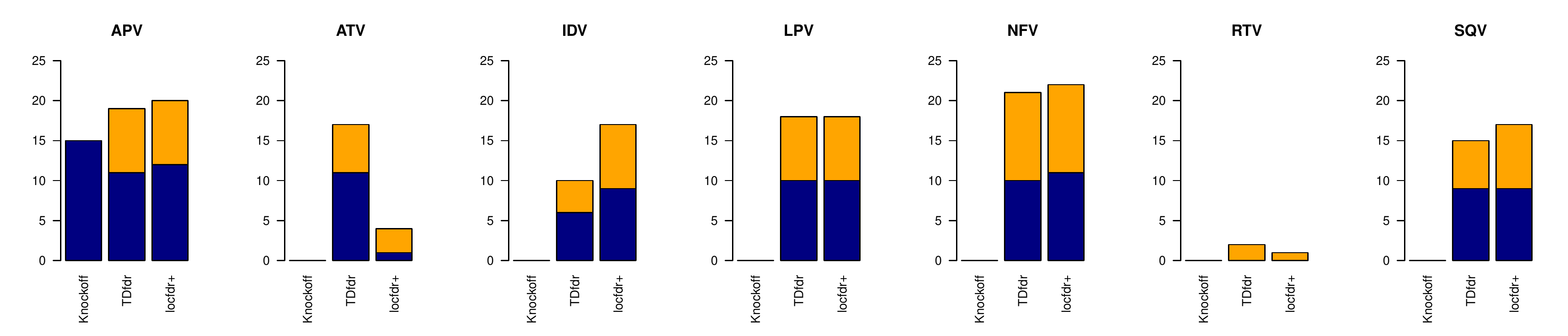}
		}
		\subfigure[FDR threshold = 10\%]{
			\includegraphics[width=1\textwidth]{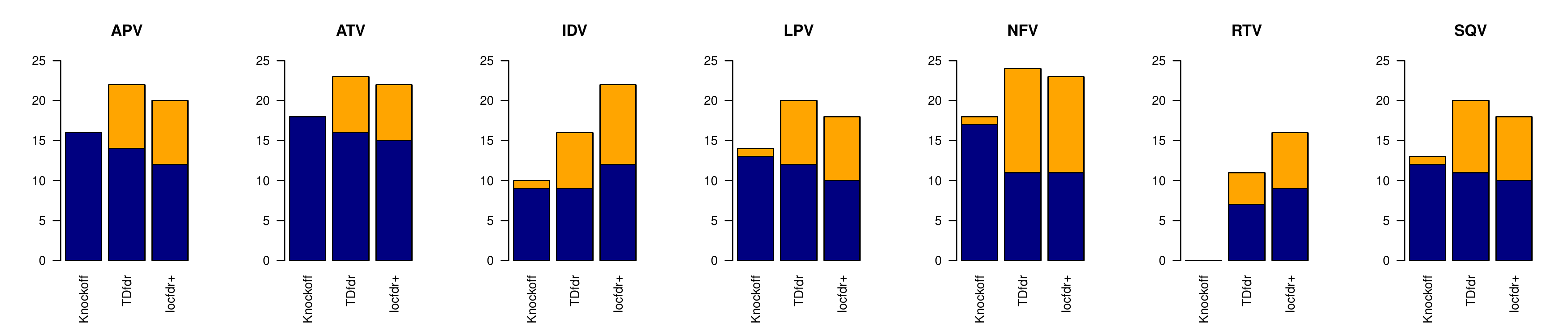}
		}
		\subfigure[FDR threshold = 20\%]{
			\includegraphics[width=1\textwidth]{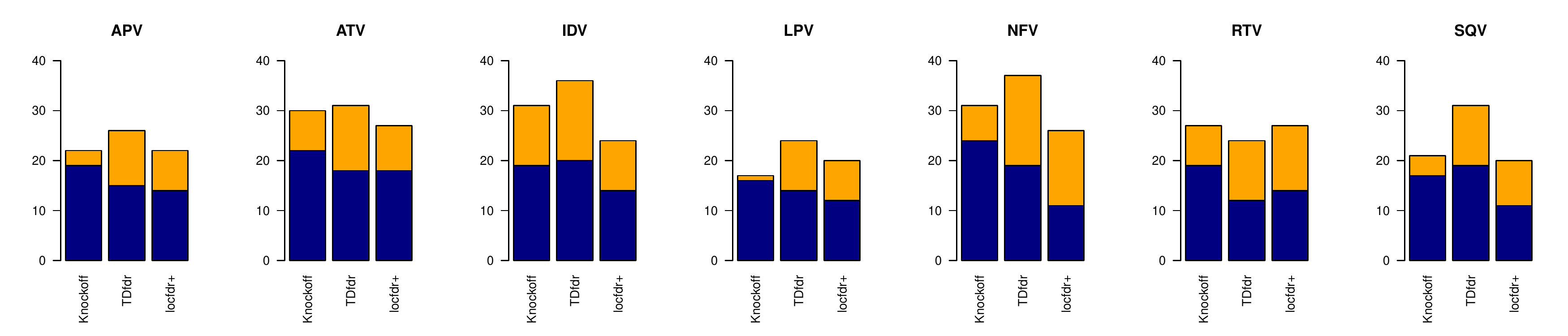}
		}
		\caption{Numbers of selected mutations by knockoff filter, TDfdr and locfdr+. The bars in blue color represent the numbers of mutations that were verified in medical assays, and the orange bars represent the mutations reported by the method yet not verified.}
		\label{fig:HIV}
	\end{figure}

	We tested TDfdr, locfdr and knockoff filter on the HIV data using Lasso scores, and reported the significant variables at different FDR thresholds of $0.01, ~0.05, ~0.1,$ and $0.2$. To be precise, the locfdr method used here was locfdr+, due to the failure of locfdr- on the Lasso scores, and the submethod for estimating the null distribution inside locfdr+ was chosen as "maximum likelihood" (nulltype=1). Seven drugs were used to test the resistance, with the names of the drugs shown as the subtitles of the plots in Figure \ref{fig:HIV}.

	In the situation where the FDR threshold is 1\%, knockoff failed to select out any variables, while TDfdr selected different numbers of variables, though some were not verified. The variables reported by locfdr+ were either empty or some unverified ones.
	As the FDR threshold increased, the powers of all methods improved as well. Knockoff filter output all-verified variables for "APV" when the FDR threshold was set as 5\%. However, for the other 6 drugs, 
	the power of knockoff filter did not improve compared to 1\%.
	For the 5\% threshold, the numbers of total selections by TDfdr all surpassed knockoff filter, and were comparable to those of locfdr+.
	More verified mutations were detected from the selection results when setting the FDR threshold at level 10\%. TDfdr still selected more mutations overall, especially for drug "RTV", for which knockoff filter failed to select any mutations. The results of TDfdr and locfdr were similar.
	As the FDR threshold increased to 20\%, all the three methods had comparable high powers.
	
	Overall, for almost all the drugs and FDR thresholds, TDfdr selected more significant mutations, including some unverified ones although their correctness is unknown.
	Knockoff filter met the serious problem of vanished power when the FDR threshold is low (1\%). This reveals a drawback of competition-based procedures, which have a "+1" correction in their FDR  estimation formula to achieve FDR control. Such "+1" correction has a side effect of dramatically decreasing the power when the number of significant variables are small and the FDR threshold is low.

	\section{Conclusion}\label{sec:conclusion}
	\setcounter{equation}{0}
	
	In this paper, we proposed an fdr estimation method, TDfdr, with and for the competition-based procedures.
	Taking the advantage of 
	competitive decoy variables,
	TDfdr provides a direct yet effective way to estimate $\pi_0$ and $f_0$. 
	Then TDfdr leverages the iteration framework of  kerfdr to estimate the fdr nonparametrically on target variables. 
	Compared to many existing methods, TDfdr extends the scope of input 
	and improves the accuracy of fdr estimation. 
	In general, the framework of TDfdr can be 
	applied to 
	any competition-based procedures, such as TDFDR and knockoff filter.
	
	Simulations on two-group data and regression data both demonstrated the higher accuracy and better stability of TDfdr than the traditional method locfdr, in terms of the averaged RMSE of estimated fdr. 
	TDfdr also estimated $\pi_0$ much more accurately than locfdr.
	Regarding FDR, 
	TDfdr is able to control the FDR under desired levels, while the locfdr method has to transform the statistics for better control of FDR, which sacrifices the power of variable selection. TDfdr, on the contrast, can give stable results, and for most of the cases TDfdr provides the most powerful selection.
	In FDR control and power, TDfdr also performed comparably or even superiorly to the TDFDR and knockoff filter methods.
	The results on the two real datasets showed the high power of TDfdr and its robustness to the FDR control levels. 
	
	In trying to apply locfdr to the scores produced by competition procedures, we met various difficulties and failures. Therefore, we do not consider locfdr an appropriate method that can be used in combination with the competition procedures. 
	
	The advantages of TDfdr over the competition procedures for FDR control lie in two aspects. First, it offers confidence assessment to individual variables. Second, it is less sensitive to small FDR control levels, overcoming the conservatism induced by the "+1" correction in FDR control methods, such as knockoff filter.
	
	In an era where high-dimensional data are  available and multiple hypothesis testing is popularly needed, TDfdr provides a new way to estimate local false discovery rate accurately with the capability to utilize any type of score. The idea of null proportion estimation can be exploited in other frameworks related to multiple hypothesis testing.

	\section*{Data Availability}
	The COVID-19 data that support the findings of this study are openly available in ProteomeXchange Consortium at \url{https://www.iprox.org/}, Project ID: IPX0002106000 and IPX0002171000.
	The HIV data that support the findings of this study are openly available in HIVDB at 
	\url{http://hivdb.stanford.edu/pages/published\_analysis/genophenoPNAS2006/}, Genotypic Predictors of Human Immunodeficiency Virus Type 1 Drug Resistance.
	
	\section*{Acknowledgements}
	
	This work was supported by the National Natural Science Foundation of China (No. 32070668).

	\bibliographystyle{unsrt} 
	\bibliography{ref}
	
		\vspace*{-0pt}

	\section*{Supplementary Information}
	
	Supplementary S1 referenced in Section~\ref{sec:localfdr}, 
	Supplementary S2-S3 referenced in Section~\ref{sec:simulation},
	and Supplementary S4 referenced in Section~\ref{sec:realdata}, 
	are available in the following  Supplementary Information.\vspace*{-70pt}


\vspace{80pt}


\renewcommand{\thefigure}{S\arabic{figure}}
\renewcommand{\thetable}{S\arabic{table}}
\renewcommand\thesection{S\arabic{section}}
\renewcommand\thesubsection{S\arabic{subsection}}
\renewcommand{\figurename}{\bf Supplementary Figure}
\captionsetup[table]{name=Supplementary Table, labelsep=period}
\setcounter {section}{0}
\setcounter {figure}{0}
\setcounter {table}{0}
\numberwithin{equation}{subsection}

\section*{\LARGE Supplementary Information}
\addcontentsline{toc}{section}{Supplementary Information}

\subsection{Competition-based procedures for FDR control} 
\label{sec:S1}

Among the existing methods for FDR control, the competition-based procedures are a new class of methods that apply to general scores other than $p$-values.
Since TDfdr is built on the competition-based procedures, here we first introduce two competition-based procedures, TDFDR\cite{He:2018}\cite{He:2018new} and knockoff filter\cite{knockoff:2015}. 
TDFDR is an approach to selecting variables that have significant differences between two groups. The knockoff filter is used for variable selection in the regression model. 
Both methods use the competitive fake variables, which are called the "decoys" in TDFDR, and "knockoffs" in the knockoff filter, to estimate the number of rejected null variables. 

As mentioned before, the problem of variable selection can be described equivalently with the language of multiple hypothesis testing. Thus we do not distinguish the terminologies in the following descriptions, e.g., we mean the same by saying "rejecting hypothesis $j$" and "selecting variable $j$ as significant".

\subsubsection{Two-group study with decoy permutations}

In the case-control two-group study, suppose we have 
$m$ variables $X_1, X_2, \ldots, X_m$ observed on $n$ subjects, $n_1$ of which are the control samples and $n_2$ ($=n-n_1$) are the case samples,
constituting the data matrix $X \in \mathbf{R}^{n \times m}$, 
with $x_{ij}$ representing the $i$-th sample of the $j$-th variable.  
Aiming at discovering the variables that are significantly different between the two groups, 
TDFDR tests the following $m$ null hypotheses,

$H_{j}$: the joint distribution of
$X_{1j},X_{2j},\ldots, X_{nj}$ is symmetric,
$j=1,2,\ldots,m$.

\noindent That is, for any random permutation $\pi$, the density function of $X_{1j},X_{2j},\ldots, X_{nj}$ satisfies $f_{X_{1j},X_{2j},\ldots, X_{nj}}(x_{1j},x_{2j},\ldots, x_{nj}) = f_{X_{1j},X_{2j},\ldots, X_{nj}}(\pi(x_{1j},x_{2j},\ldots, x_{nj}))$.

To measure the difference of a variable between two groups, a scoring function $s$ satisfying $s\left(x_{1j}, \cdots, x_{nj}\right)=s\left(\pi\left(x_{1j}, \cdots, x_{n_{1},j}\right), \pi\left(x_{n_{1}+1,j}, \cdots, x_{nj}\right)\right)$ is needed, such as $t$-test statistic.
Without loss of generality, we assume that larger values of scores represent larger differences between groups.

For each variable $j$, TDFDR first calculates an "original score" $S_{j}^o = s\left(x_{1j}, \cdots, x_{nj}\right)$ for the original sample and $N$ "permutation scores" $S_{jk}^p = s\left(\pi_k\left(x_{1j}, \cdots, x_{nj}\right)\right), k=1,2,\ldots,N$ on the permuted samples  
and then sorts the $N+1$ scores in descending order.

Next, TDFDR assigns each variable a label of "target" or "decoy", by comparing the original score with a statistic of the $N$ permutation scores, where the statistic can be maximum, median, or other statistics representing the population characteristic of the permutation scores.
Here we utilize the "median" rule to label variable $j$ as follows:

$$	L_{j}=\left\{\begin{array}{ll}
		{T,} & {R_j<(N+2) / 2} \\
		{T/D,} & {R_j=(N+2) / 2} \\
		{D,} & {R_j>(N+2) / 2}
	\end{array}\right.  $$
where $R_j$ is the rank of the original score in the $N+1$ scores, $T$ and $D$ represent "target" and "decoy" respectively.
Note that if $N$ is an even number, there exists the possibility of $R_j$ satisfying $R_j=(N+2)/2$, and we label the corresponding variable randomly as "target" or "decoy" with equal probability, i.e., $P(L_j=T)=P(L_j=D)=\frac{1}{2}$.
This labelling creates a division to the variables. We define $\mathcal{T} := \{j=1,2,\ldots,m:L_j=T\}$ and $\mathcal{D} := \{j=1,2,\ldots,m:L_j=D\}$.

After each variable is assigned a label, a final score is determined for it: 
$$	S_{j}=\left\{\begin{array}{ll}
		{S_{j}^{o},} & {j \in \mathcal{T}} \\
		{\tilde{S}_{j}^{(N+2-R_j)},} & {j \in \mathcal{D}}
	\end{array}\right.  $$
where $\tilde{S_j}$ is the sorted $N+1$ scores.  When the label is target, we set the original score as the final score directly; when the label is decoy,
the final score is set as the permutation score ranking at the symmetric position of $R_j$ about the median score. 

There exist other ways to label and score the variables, which could enhance the power \cite{He:2018}\cite{emery:2021}\cite{emery:thesis}.

The variables are then sorted according to their final scores decreasingly, and the label of the $j$-th variable in the sorted list is denoted by $L_{(j)}$. With this notation, a higher-scored variable with $L_{(j)} = T$ potentially has a less possibility of being a true null.

Finally, those variables with $L_{(j)} = T$ and $j \le K_{td}$ are selected, where $K_{td}$ is determined by 
\begin{equation}\label{TDFDRfilt}
	K_{td}=\max \left\{k = 1,\ldots,m : \frac{\#\left\{j \leq k:L_{(j)}=D \right\}+1}{\#\left\{j \leq k:L_{(j)}=T\right\} \vee 1} \leq q\right\}
\end{equation}
The selected variables are considered as significantly different between the case and control groups. 
It can be proven that the TDFDR procedure controls the FDR under level $q$.

Note that the proof of FDR control of the TDFDR method relies on the assumption of independence between variables, though practically the TDFDR method shows good control of FDR for data with dependency.

\subsubsection{Variable selection with knockoffs}\label{sec:knockoff}

As mentioned before, 
knockoff filter\cite{knockoff:2015} is an influential 
competition-based method for FDR control, mainly used in the context of linear regression model:
$$
	\mathbf{y}=\mathbf{X} \boldsymbol{\beta}+\mathbf{z}.
	$$
where $\mathbf{y} \in \mathbb{R}^{n}$ is a vector of responses, $\mathbf{X} \in \mathbb{R}^{n \times m}$ is the  design matrix, $\boldsymbol{\beta} \in \mathbb{R}^{m}$ is an unknown vector of coefficients and $\mathbf{z} \sim \mathcal{N}\left(0, \sigma^{2} \mathbf{I}\right)$ is Gaussian noise. The aim of the regression model is to search for the variables whose coefficients are nonzero.

The knockoff method first constructs "knockoff" matrix $\tilde{\mathbf{X}}$ so that it exhibits the same covariance structure as the original design matrix, but in addition, the correlations between distinct original and knockoff variables are the same as those between the distinct variables in the original matrix. 
Let $\mathbf{\Sigma} = \mathbf{X}^{\top}\mathbf{X}$ be the Gram matrix.
By requiring the "knockoff" matrix obey that
$$	\tilde{\mathbf{X}}^{\top} \tilde{\mathbf{X}}=\mathbf{\Sigma}, \quad \mathbf{X}^{\top} \tilde{\mathbf{X}}=\mathbf{\Sigma}-{diag}\{\mathbf{s}\},  $$
the knockoff matrix can be solved as
$$	\tilde{\mathbf{X}}=\mathbf{X}\left(\mathbf{I}-\mathbf{\Sigma}^{-1} {diag}\{\mathbf{s}\}\right)+\tilde{\mathbf{U}} \mathbf{C},  $$
where $\mathbf{s}$ is an $m$-dimensional nonnegative vector, $\tilde{\mathbf{U}}$ is an $n \times m$ orthonormal matrix that is orthogonal to the span of the design matrix
$\mathbf{X}$, and $\mathbf{C}^{\top} \mathbf{C}=2 {diag}\{\mathbf{s}\}-{diag}\{\mathbf{s}\} \mathbf{\Sigma}^{-1} {diag}\{\mathbf{s}\} \succeq \mathbf{0} $ is a Cholesky decomposition.
Through maximizing the diagonal entries in $\mathbf{s}$, the knockoff filter makes the correlations between the knockoff variables and the true signals as small as possible.
Note that the construction above is only suited for the situation where $n \ge 2m$. The knockoff filter can also be extended to $m < n < 2m$ with certain settings. 

A statistic $Z_j$ can be computed to measure the relevance of the original variable $X_j$ to the response variable, 
and similarly, $\tilde{Z_j}$ for the knockoff variable. 
For instance, in the Lasso model, the statistics $(Z_1,\ldots,Z_m,\tilde{Z_1}, \ldots,\tilde{Z}_m)$ can be computed by solving the optimization $\hat{\boldsymbol{\beta}}(\lambda)=\underset{\boldsymbol{\beta}}{\operatorname{argmin}}\left\{\frac{1}{2}\|\mathbf{y}-[\mathbf{X}~\tilde{\mathbf{X}}] \boldsymbol{\beta}\|_{2}^{2}+\lambda\|\boldsymbol{\beta}\|_{1}\right\}$, with $Z_j$ ($\tilde{Z_j}$) representing the largest value of penalty tuning parameter $\lambda$ when variable $X_j$ ($\tilde{X_j}$) enters the Lasso path. Note that the design matrix in the general Lasso model is replaced by $[\mathbf{X}~ \tilde{\mathbf{X}}]$ (the columnwise concatenation of $\mathbf{X}$ and $\mathbf{\tilde{X}}$) to achieve competition between variables, and the length of $\boldsymbol{\beta}$ is also doubled. 

In order to tease apart those variables that are in the regression model (i.e. $\beta_j \ne 0$) from those that are not (i.e. $\beta_j = 0$), test statistics $W_j$'s are constructed so that large positive values are evidence against the null hypothesis $\beta_j = 0$. 
As long as a statistic satisfies the sufficiency property (i.e., $W_j$ depends only on the Gram matrix and variable-response inner products) and the antisymmetry property (i.e., swapping $X_j$ and $\tilde{X_j}$ has the effect of switching the sign of $W_j$), it can be chosen as a proper test statistic for the knockoff method.
For instance, 
$$	W_{j}=Z_{j} \vee \tilde{Z}_{j} \cdot\left\{\begin{array}{ll}
		+1, & Z_{j}>\tilde{Z}_{j}, \\
		-1, & Z_{j}<\tilde{Z}_{j}, \\
		0~, & Z_{j}=\tilde{Z}_{j}
	\end{array}\right.  $$
Other forms of statistics that satisfy the two properties can also be chosen depending on the circumstances. In fact, the principle in the antisymmetry property is that $W_j$ is yielded by competition between $X_j$ and $\tilde{X_j}$.

Finally, variables are selected with $W_j > T_{ko}$ , where $T_{ko}$ is determined as
\begin{equation}\label{knockofffilt}
	T_{ko}=\min \left\{t \in \mathcal{W}: \frac{\#\left\{j: W_{j} \leq-t\right\}+1}{\#\left\{j: W_{j} \geq t\right\} \vee 1} \leq q \right\}
\end{equation}
where $q$ is the FDR control level and $\mathcal{W}=\left\{\left|W_{j}\right|: j=1, \ldots, m\right\} \backslash\{0\}$ is the set of unique nonzero values of $|W_j|$’s.

It is proven that, with the exchangeability property of statistics $W_j$'s, the knockoff method is able to control FDR at the given threshold $q$ under arbitrary variable dependency.

\subsubsection{The connection of TDFDR and knockoff filter}\label{sec:test}

	Note that the symmetric rule of $W_j$'s in the knockoff filter is equivalent to giving the variables a division, by the signs of $W_j$'s. 
	Therefore, the knockoff procedure can be described in the target-decoy framework, by replacing the signs of statistics $W_j$'s with labels of "target" or "decoy" and defining the absolute values of $W_j$'s as the final scores, and vice versa.
	
	On the whole, both TDFDR and knockoff filter are competition-based procedures. 
	First, they both create new "fake"  variables which are called decoys or knockoffs. 
	Second, the "fake" variables compete with their corresponding original variables to produce antisymmetric ranking statistics. That is, for true nulls, their statistics have equal probability of being target (positive) or decoy (negative). 
	Third,
	they use the same formula 
	to compute the rejection region, i.e., Equation S1.1
	and 
	S1.2. Note that there is a "+1" term in both equations, which is essential for FDR control, and was first proposed in the context of mass spectrometry based proteomics \cite{He:2015}\cite{He:2013}.

\vspace{50pt}
	
	\subsection{Simulation design}
	
	\subsubsection{Two-group study}
	For the basic two-sample simulation, we chose two scenarios, where the data were sampled from normal or gamma distributions.

	In the normal scenario, we sampled all the control data from $N(0, 1)$;
	on the contrary, the case data were sampled from $N(0, 1)$ for the null hypotheses and $N(a, 1)$ for non-null hypotheses, respectively, where the mean  parameter $a$ controls the difference between the two groups. We simulated different configurations of $a=2, 2.5, 3$. 
	In the gamma scenario, we sampled all the control data from $Ga(2, 1)$; the case data were sampled from $Ga(2, 1)$ for null hypotheses and
	$Ga(\alpha, 1)$ for non-null hypotheses, respectively.
	In the simulations, $\alpha$ was set as $6,7,8$. 
	For other parameters in both normal and gamma scenarios, we chose the number of hypotheses $m = 10000$, i.e., each  sample contained $10000$ variables; the sample size of each control/case group $g = 5$; the proportion of null hypotheses $\pi_0 = 0.8, 0.9, 0.95$. 
	After the generation of random samples, the $t$-test statistic was used as the scoring function to characterize the differences between the case and control groups. 
	In this way, the higher a score is, the larger difference it represents between the two groups.
	
	The number of permutations in TDfdr was set as $N = 19$. For locfdr, the parameter of estimating null distribution was chosen to be "maximum likelihood" (nulltype=1, the default). The transformation of the $t$-statistic $S$ in the two-group study is 
		$	S^{'} = \Phi^{-1}(\psi(S))  $
	, where $\Phi$ and $\psi$ are the cumulative density function of the standard normal distribution and the probability density function of the $t$ distribution with a prespecified degree of freedom, respectively.
	Note that in the Equation \ref{equ:fdrRMSE} there is a varying set of interest $\mathcal{I}$, so here in the two-group simulations, $\mathcal{I}^{(k)}=\mathcal{T}^{(k)}$ for TDfdr, and $\mathcal{I}^{(k)} = \{1,\dots,m\}$ for locfdr- and locfdr+, for $k=1,2,\ldots,M$.

	In FDR evaluation, we used the TDFDR\cite{He:2018new} method for comparison.

	\subsubsection{Regression model}
	
	We used the regression model described in 
	Supplementary S1.2 
	and simulation settings of the knockoff filter\cite{knockoff:2015} to test the effect of TDfdr.
	
	First, the design matrix was generated row by row i.i.d. from an $\mathcal{N}(\mathbf{0}, \mathbf{\Theta})$ distribution, where 
	$\Theta_{ij}=\rho^{|i-j|}$ for 
	$i,j=1,\ldots,m$. Then we centered and normalized the columns of $\mathbf{X}$ and calculated the simulation value of $\mathbf{y}$ as
	$\mathbf{y}=\beta \cdot\left(\mathbf{X}_{1}+\cdots+\mathbf{X}_{m_1}\right)+\mathbf{z} $
	, where $\mathbf{z} \sim \mathcal{N}\left(\mathbf{0}, \mathbf{I}_{n}\right)$, and $m_1$ is the number of significant variables among all $m$. Thus, the null proportion for these $m$ hypotheses is $\pi_0 = 1- m_1/m$.
	
	We simulated $m=2000$ variables in the regression model, in which $m_1=100, 200, 400$ variables were significant, 
	i.e., $\pi_0=0.95, 0.9, 0.8$. For each variable, 6000 samples were simulated from the multiple normal distribution. Besides, 
	we simulated two cases of variables with and without dependency,  corresponding to the correlation coefficient $\rho=0$ and $\rho=0.3$. 
	To vary the difficulty of variable selection, we sampled $\beta_j$ randomly from $\{\pm A\}$ for each of the $m_1$ selected coefficients, where the signal amplitude $A = 2.5, 3.5, 4.5$.

	Fitting the data in the Lasso model with the  concatenated design matrix from original and knockoff ones, 
	we computed the statistics $W_j$'s as 
	in the example of Supplementary S1.2.
	For the knockoff-based simulation, fdr can be estimated as in the two-group study, 
	yet using the Lasso statistic $W_j$'s.
			The locfdr method was also used to estimate the fdr for comparison. 
			Locfdr- took the original Lasso statistics as input, but failed to complete valid fdr estimation due to the violation of normal assumption of inputs. In carrying out locfdr+, there was also an obstacle when transforming the Lasso statistics, because the theoretical null distribution of them is unknown. 
			Finally, we leveraged the decoy/knockoff variables 
			to estimate an empirical null, then transformed the remaining "target" statistics using it. 
			Formally, the transformation for a Lasso statistic $S$ in the regression model is 
				$	S^{'} = \Phi^{-1}(\Psi(S)),$
			where $\Phi$ and $\Psi$ are the cumulative density function of the standard normal distribution and the empirical cumulative density function estimated from the decoy variables, respectively.
			Thus the locfdr+ estimates the fdr of "target" variables here. 
			As a result, we compared TDfdr and locfdr+ in terms of the accuracies of $\pi_{0}$ and fdr estimation, with $\mathcal{I}^{(k)}=\mathcal{T}^{(k)}, ~k=1,2,\ldots,M$ in Equation \ref{equ:fdrRMSE} for both methods. To be precise, here the $\pi_0$ is in fact $\pi_{0t}$.
			The remaining parameters such as "permutation time" in TDfdr and the "nulltype" in locfdr were set as the same as those in the two-group simulations.
			
			FDR control and power are also evaluated, with the knockoff filter as the benchmark. The computation from fdr to FDR was the same as that in two-group simulation.

		\newpage
			
	\subsection{Simulation results}
	\subsubsection{Two-group study}\label{sec:simuresults-twogroup}
	~\\
	
	\begin{figure}[H]
		\centering
		\subfigure[$\pi_0=0.8$]{
			\begin{minipage}[t]{\linewidth}
				\centering
				\includegraphics[width=1\textwidth, height=5.5cm]{figures/normal-fdrRMSE08-410-RMSE.pdf}
			\end{minipage}
		}%
		\\
		\subfigure[$\pi_0=0.9$]{
			\begin{minipage}[t]{\linewidth}
				\centering
				\includegraphics[width=1\textwidth, height=5.5cm]{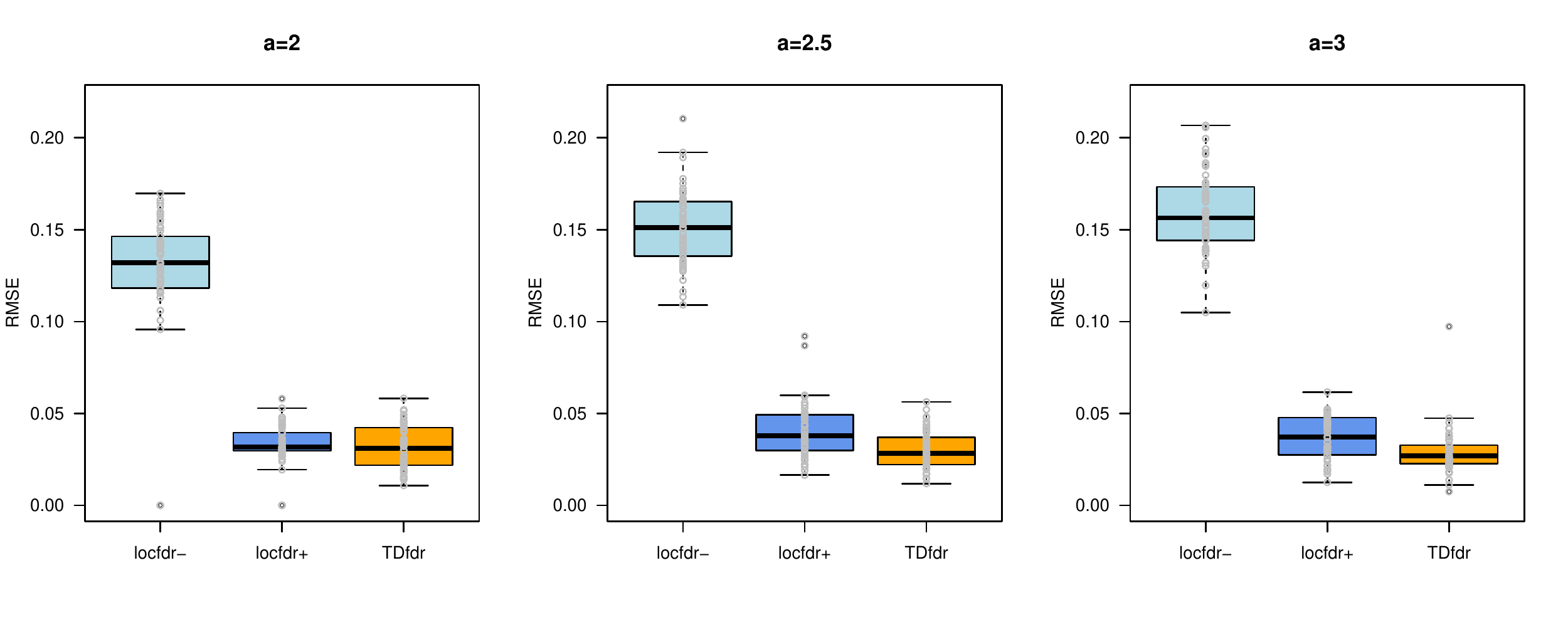}
			\end{minipage}
		}
		\\
		\subfigure[$\pi_0=0.95$]{
			\begin{minipage}[t]{\linewidth}
				\centering
				\includegraphics[width=1\textwidth, height=5.5cm]{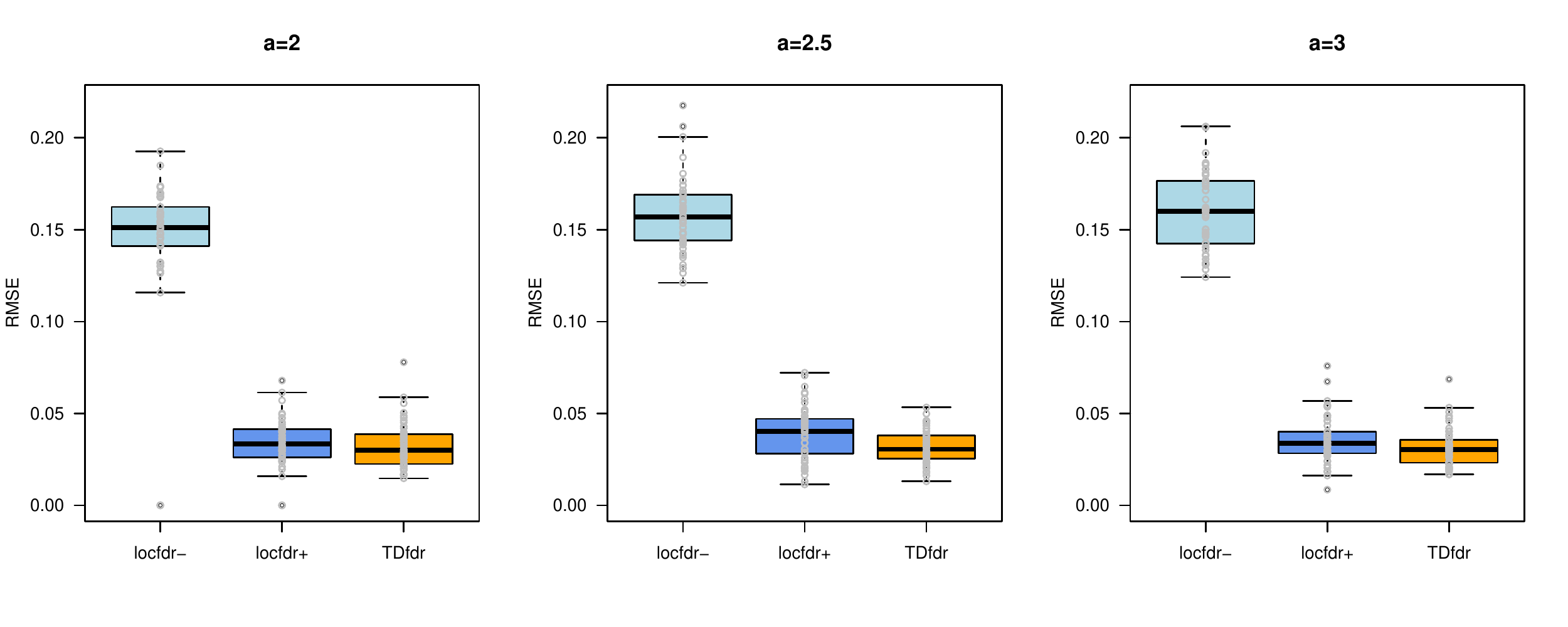}
			\end{minipage}
		}
		\caption{RMSEs of fdr estimation of normal data}
		\label{fig:normalSimu-fdr}
	\end{figure}

	\begin{figure}[H]
		\centering
		\subfigure[$\pi_0=0.8$]{
			\begin{minipage}[t]{\linewidth}
				\centering
				\includegraphics[width=1\textwidth, height=5.5cm]{figures/gamma-fdrRMSE08-410-RMSE.pdf}
			\end{minipage}%
		}%
		\\
		\subfigure[$\pi_0=0.9$]{
			\begin{minipage}[t]{\linewidth}
				\centering
				\includegraphics[width=1\textwidth, height=5.5cm]{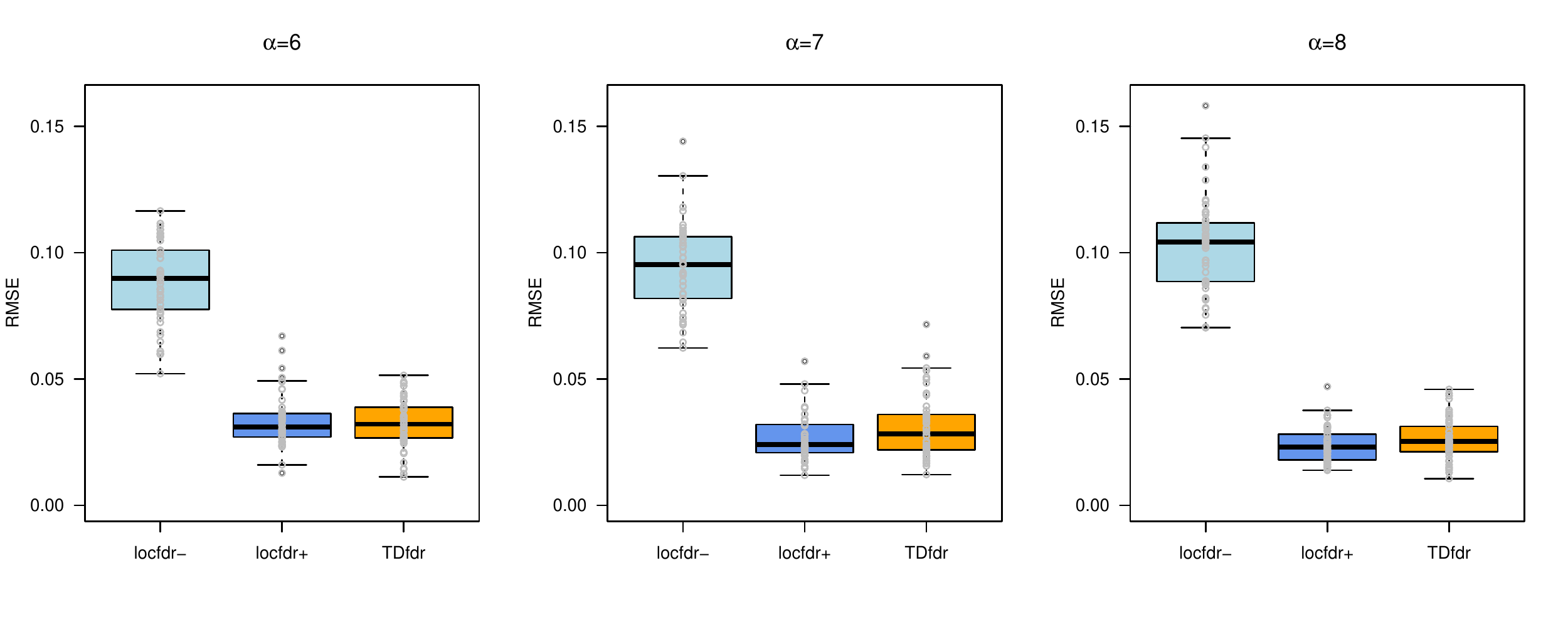}
			\end{minipage}
		}
		\\
		\subfigure[$\pi_0=0.95$]{
			\begin{minipage}[t]{\linewidth}
				\centering
				\includegraphics[width=1\textwidth, height=5.5cm]{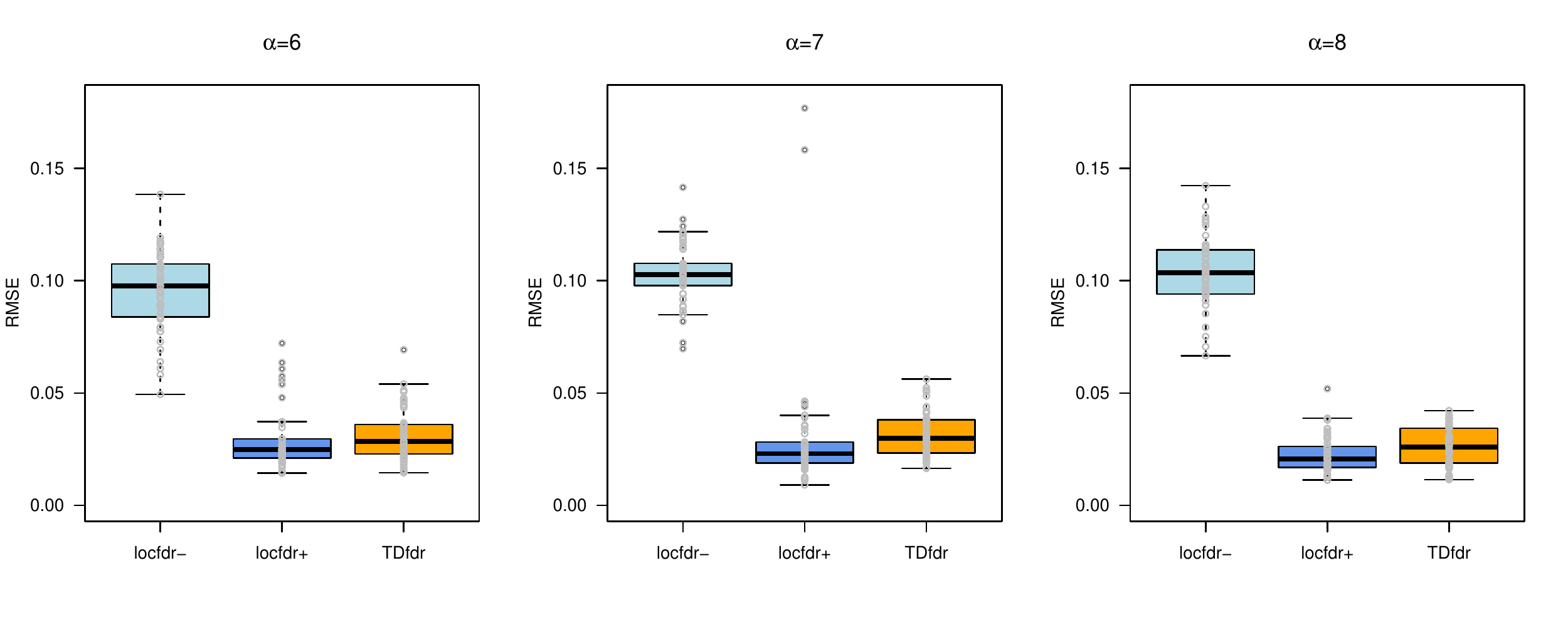}
			\end{minipage}
		}
		\caption{RMSEs of fdr estimation of gamma data}
		\label{fig:gammaSimu-fdr}
	\end{figure}

	\begin{figure}[H]
		\setcounter{subfigure}{0}
		\centering
		\subfigure[$\pi_0=0.8$]{
			\begin{minipage}[t]{\linewidth}
				\centering
				\includegraphics[width=1\textwidth]{figures/normal-control08-39.pdf}
			\end{minipage}%
		}%
		\\
		\subfigure[$\pi_0=0.9$]{
			\begin{minipage}[t]{\linewidth}
				\centering
				\includegraphics[width=1\textwidth]{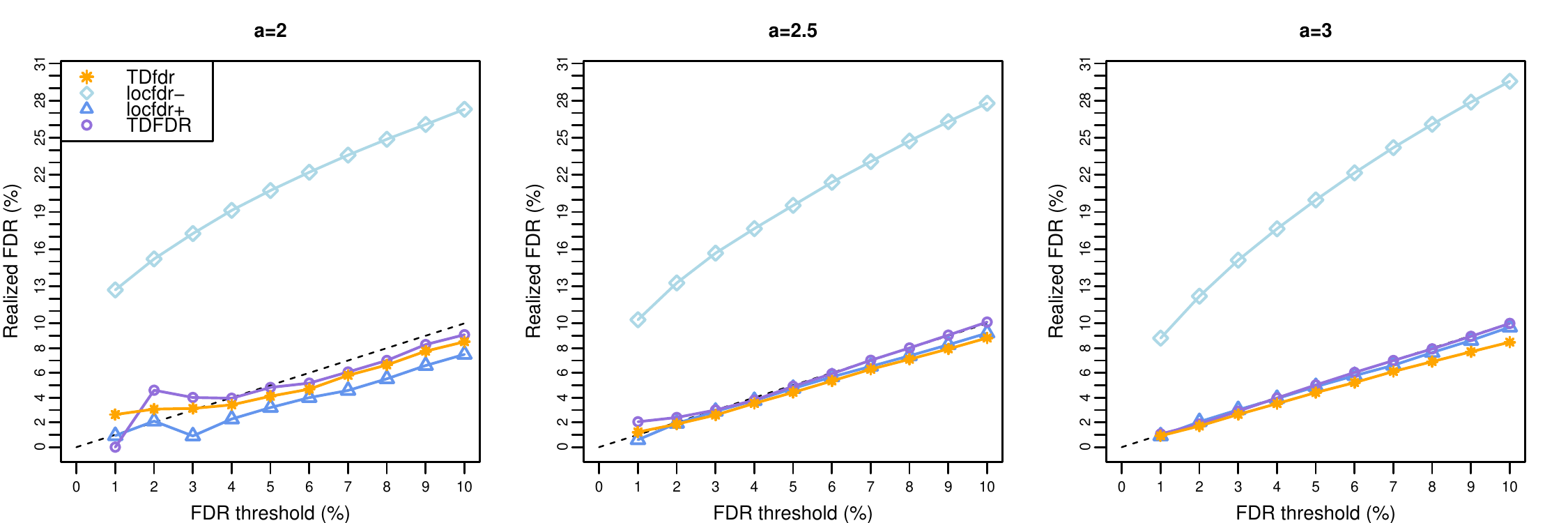}
			\end{minipage}%
		}%
		\\
		\subfigure[$\pi_0=0.95$]{
			\begin{minipage}[t]{\linewidth}
				\centering
				\includegraphics[width=1\textwidth]{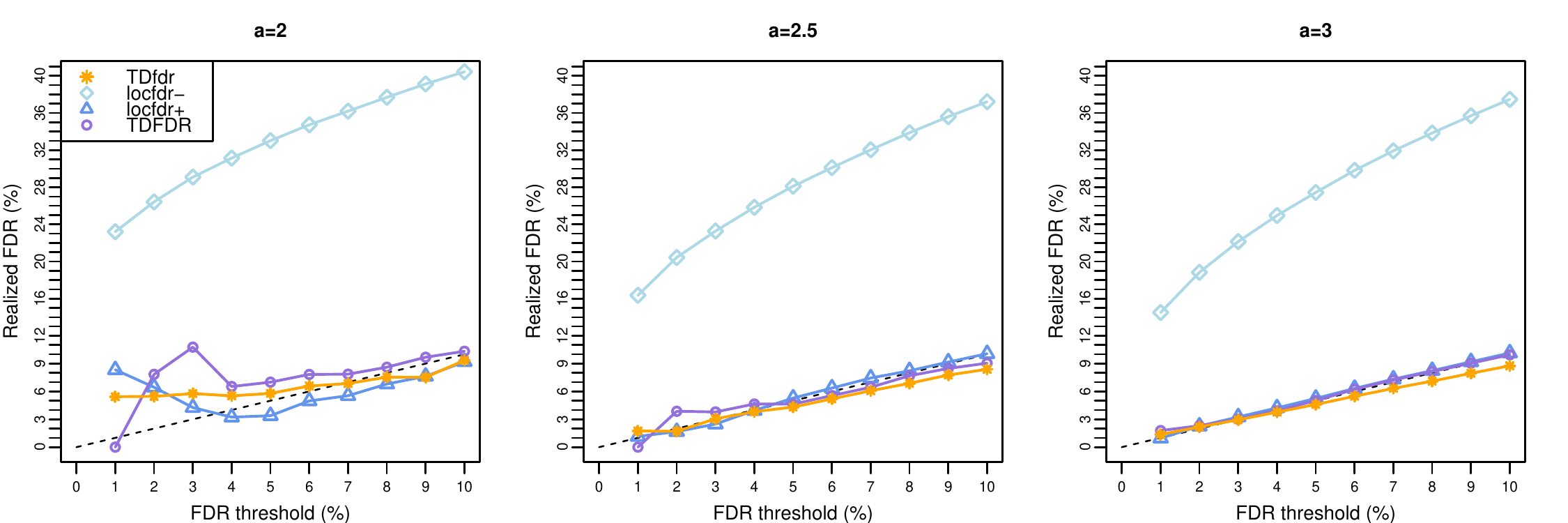}
			\end{minipage}
		}
		\caption{FDR control results of normal data}
		\label{fig:normalSimu-control}
	\end{figure}

	\begin{figure}[H]
		\setcounter{subfigure}{0}
		\centering
		\subfigure[$\pi_0=0.8$]{
			\begin{minipage}[t]{\linewidth}
				\centering
				\includegraphics[width=1\textwidth]{figures/gamma-control08-39.pdf}
			\end{minipage}%
		}%
		\\
		\subfigure[$\pi_0=0.9$]{
			\begin{minipage}[t]{\linewidth}
				\centering
				\includegraphics[width=1\textwidth]{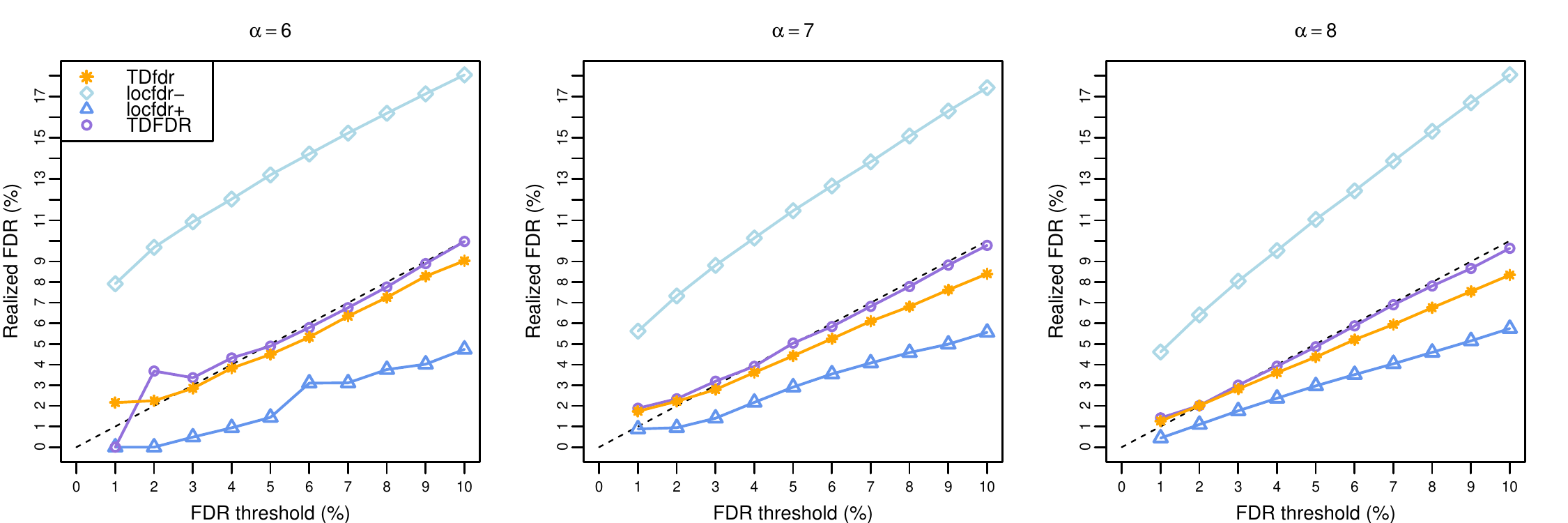}
			\end{minipage}%
		}%
		\\
		\subfigure[$\pi_0=0.95$]{
			\begin{minipage}[t]{\linewidth}
				\centering
				\includegraphics[width=1\textwidth]{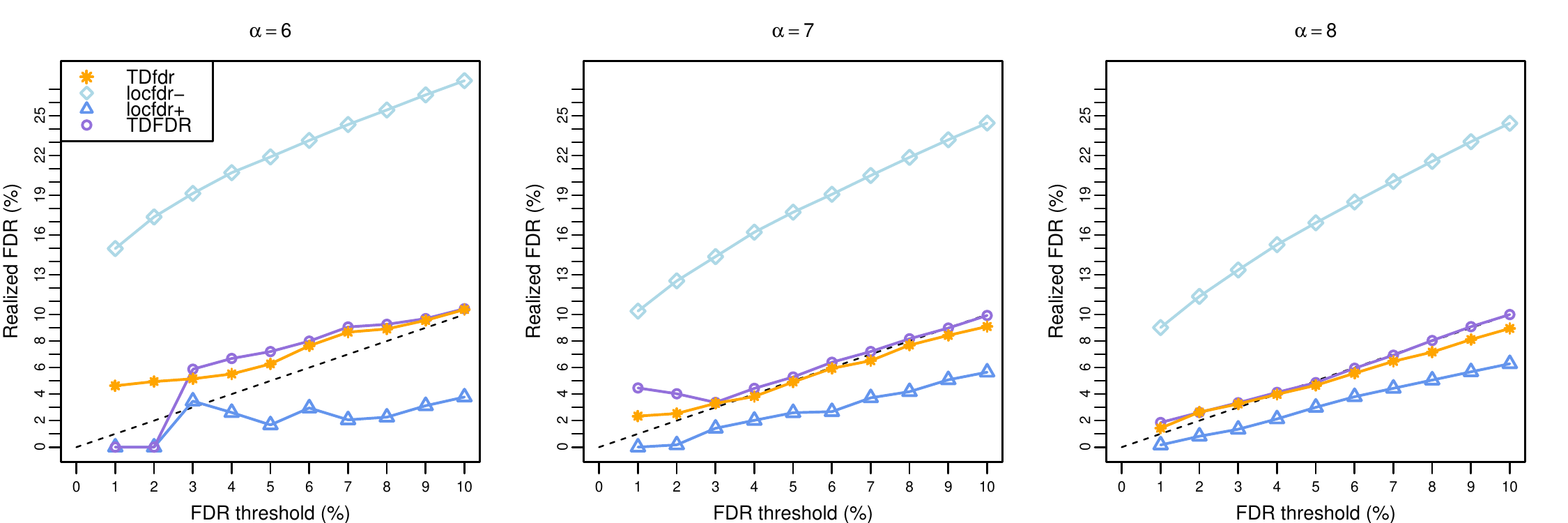}
			\end{minipage}
		}
		\caption{FDR control results of gamma data}
		\label{fig:gammaSimu-control}
	\end{figure}

	\begin{figure}[H]
		\setcounter{subfigure}{0}
		\centering
		\subfigure[$\pi_0=0.8$]{
			\begin{minipage}[t]{\linewidth}
				\centering
				\includegraphics[width=\textwidth]{figures/normal-power08-312.pdf}
			\end{minipage}%
		}%
		\\
		\subfigure[$\pi_0=0.9$]{
			\begin{minipage}[t]{\linewidth}
				\centering
				\includegraphics[width=\textwidth]{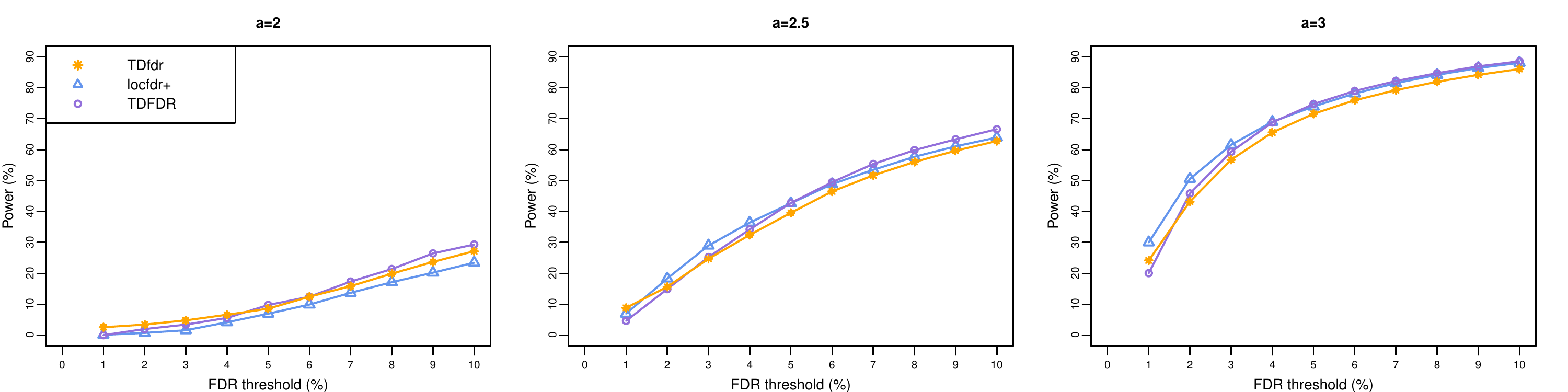}
			\end{minipage}%
		}%
		\\
		\subfigure[$\pi_0=0.95$]{
			\begin{minipage}[t]{\linewidth}
				\centering
				\includegraphics[width=\textwidth]{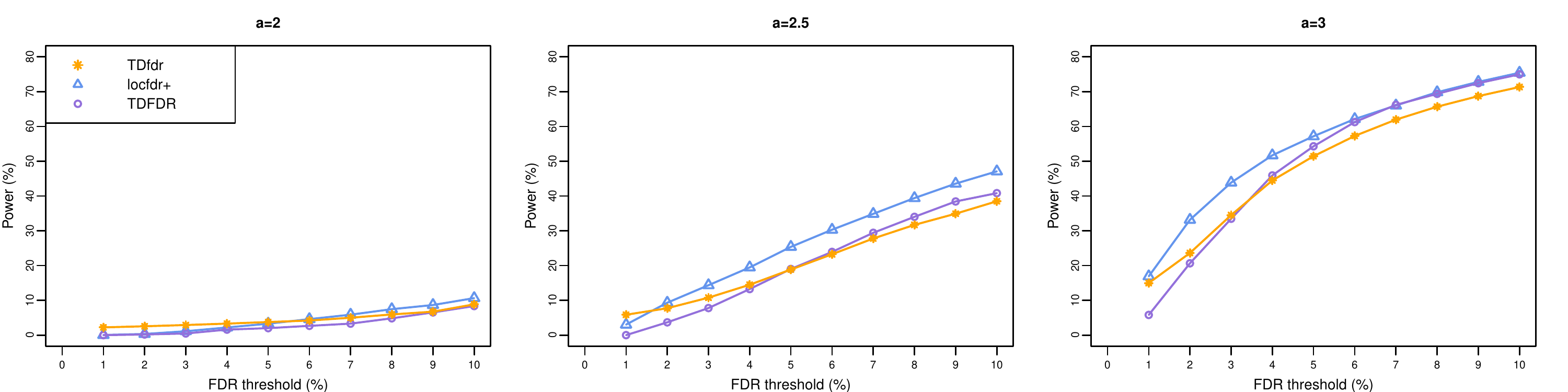}
			\end{minipage}%
		}
		\centering
		\caption{Powers of normal data}
		\label{fig:normalSimu-power}
	\end{figure}

	\begin{figure}[H]
		\setcounter{subfigure}{0}
		\centering
		\subfigure[$\pi_0=0.8$]{
			\begin{minipage}[t]{\linewidth}
				\centering
				\includegraphics[width=\textwidth]{figures/gamma-power08-312.pdf}
			\end{minipage}
		}
		\\
		\subfigure[$\pi_0=0.9$]{
			\begin{minipage}[t]{\linewidth}
				\centering
				\includegraphics[width=\textwidth]{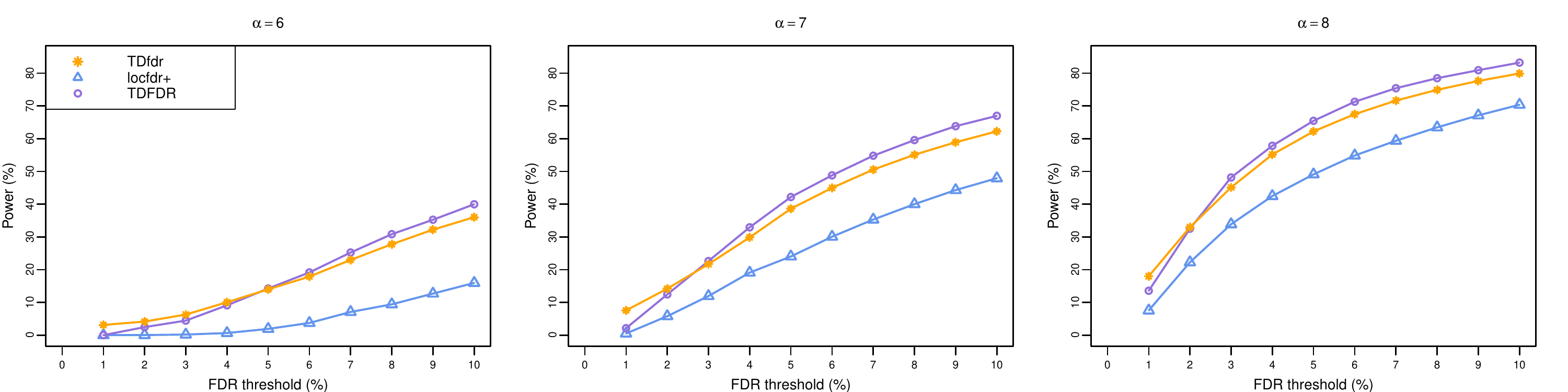}
			\end{minipage}%
		}%
		\\
		\subfigure[$\pi_0=0.95$]{
			\begin{minipage}[t]{\linewidth}
				\centering
				\includegraphics[width=\textwidth]{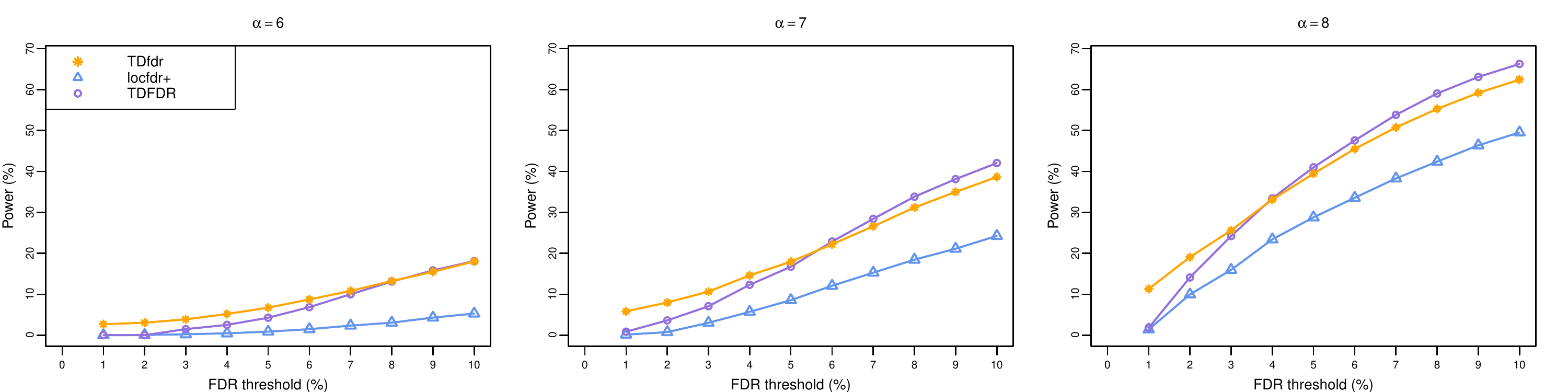}
			\end{minipage}%
		}
		\centering
		\caption{Powers of gamma data}
		\label{fig:gammaSimu-power}
	\end{figure}

\vspace{200pt}

	\subsubsection{Regression model}\label{sec:simuresults-regression}
	~\\
	
	\begin{figure}[H]
		\centering
		\subfigure[$\pi_0 = 0.8$]{
			\begin{minipage}[t]{1\linewidth}
				\centering
				\includegraphics[width=\textwidth]{figures/KOfdr-08-39-RMSE.pdf}
			\end{minipage}%
		}
		\\
		\subfigure[$\pi_0 = 0.9$]{
			\begin{minipage}[t]{1\linewidth}
				\centering
				\includegraphics[width=\textwidth]{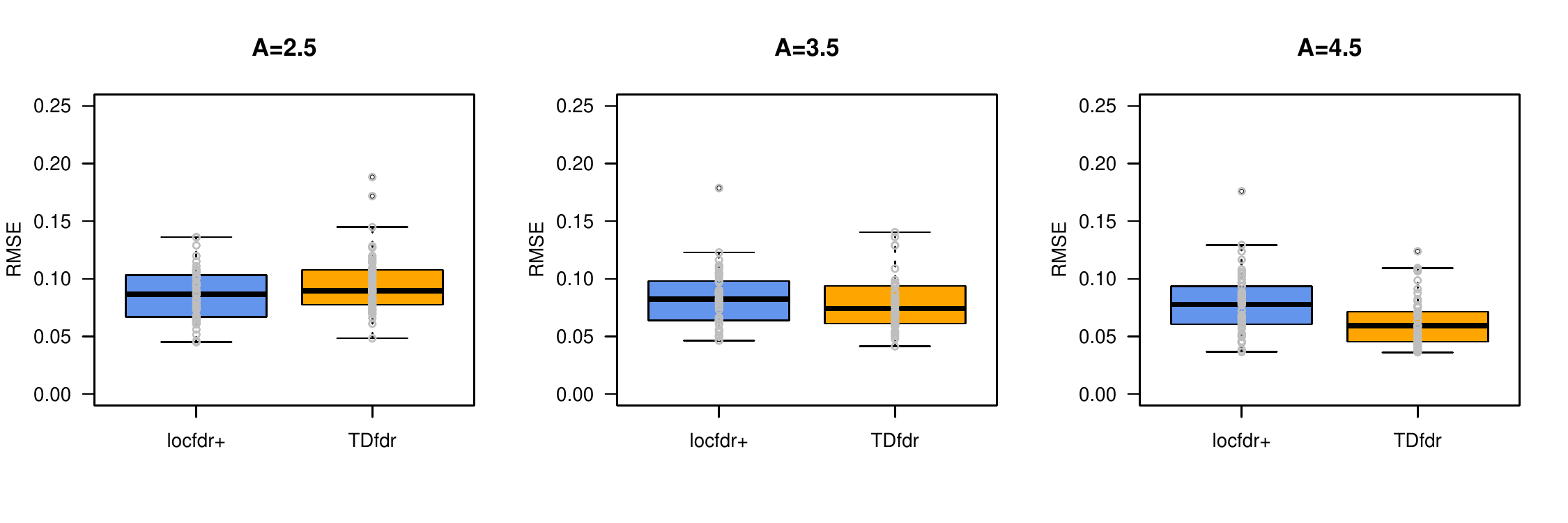}
			\end{minipage}%
		}
		\\
		\subfigure[$\pi_0 = 0.95$]{
			\begin{minipage}[t]{1\linewidth}
				\centering
				\includegraphics[width=\textwidth]{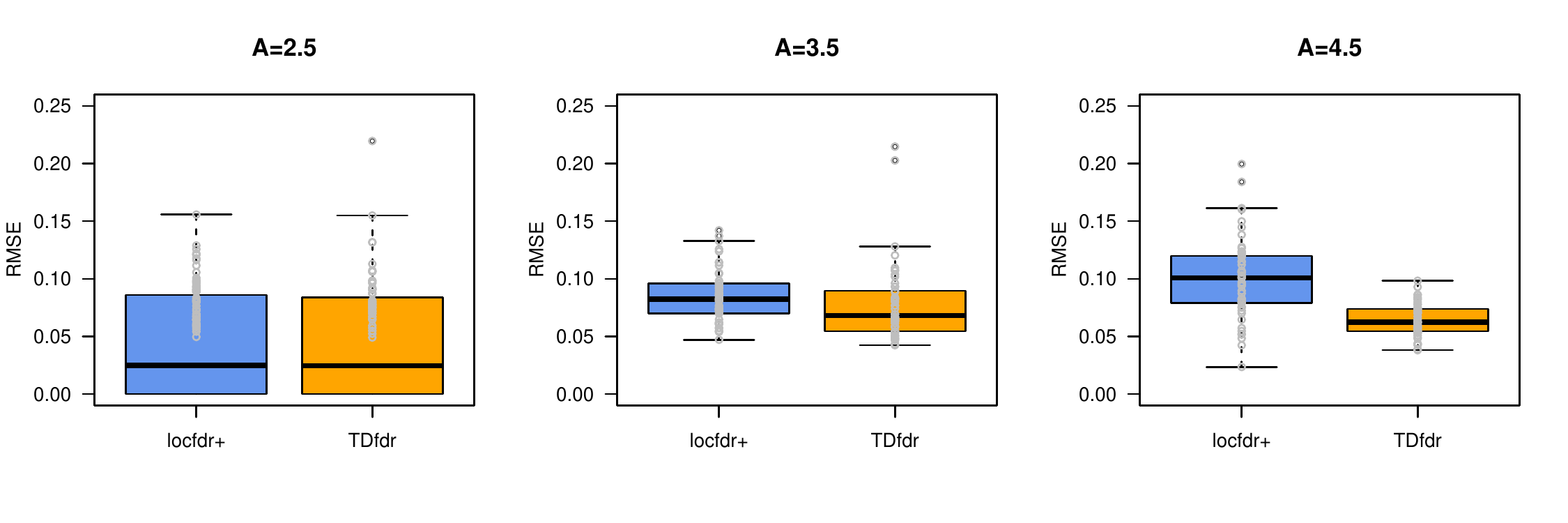}
			\end{minipage}%
		}%
		\centering
		\caption{RMSEs of fdr estimation of  regression data (independent cases)}
		\label{fig:koSimu-fdr}
	\end{figure}

	\begin{figure}[H]
		\centering
		\subfigure[$\pi_0 = 0.8$]{
			\begin{minipage}[t]{1\linewidth}
				\centering
				\includegraphics[width=\textwidth]{figures/KOfdr-08-39-corr-RMSE.pdf}
			\end{minipage}%
		}
		\\
		\subfigure[$\pi_0 = 0.9$]{
			\begin{minipage}[t]{1\linewidth}
				\centering
				\includegraphics[width=\textwidth]{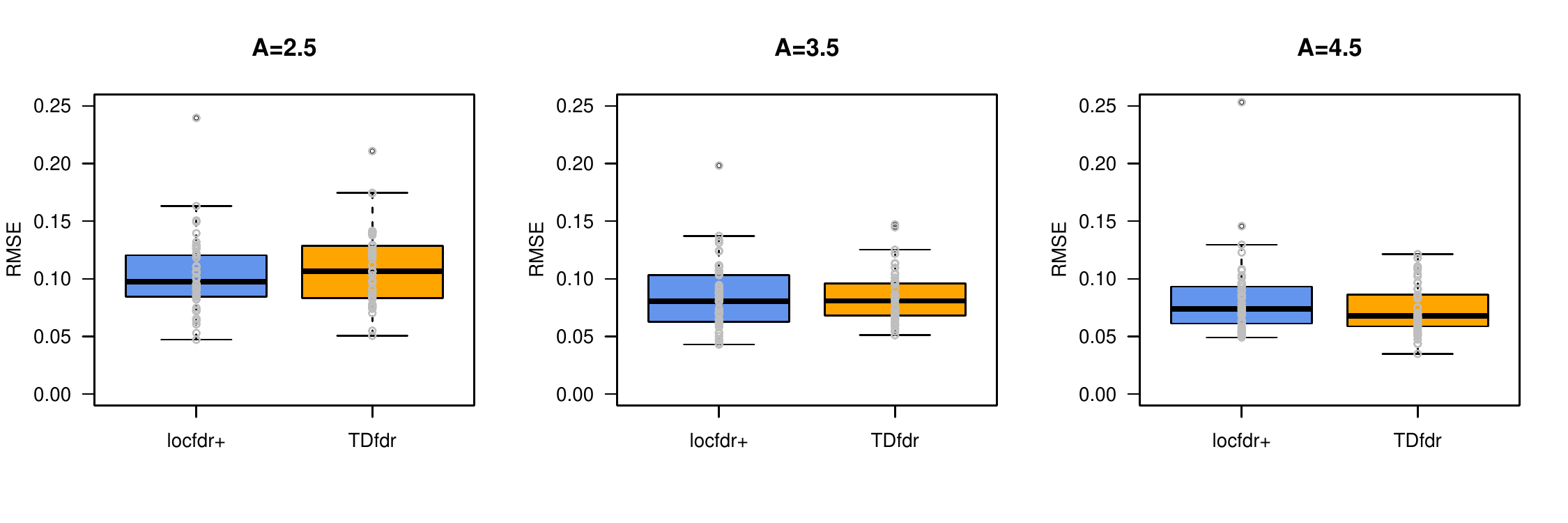}
			\end{minipage}%
		}%
		\\
		\subfigure[$\pi_0 = 0.95$]{
			\begin{minipage}[t]{1\linewidth}
				\centering
				\includegraphics[width=\textwidth]{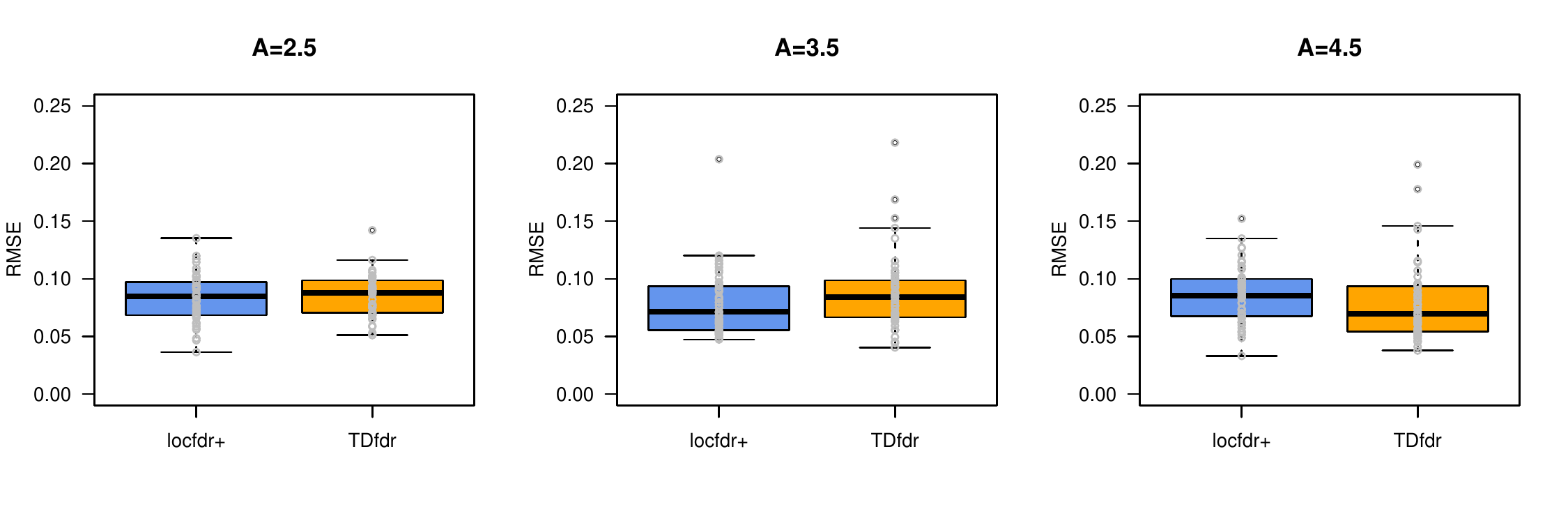}
			\end{minipage}%
		}%
		\centering
		\caption{RMSEs of fdr estimation of  regression data (dependent cases)}
		\label{fig:koSimu-fdr-corr}
	\end{figure}

	\begin{figure}[H]
		\centering
		\subfigure[$\pi_0 = 0.8$]{
			\begin{minipage}[t]{1\linewidth}
				\centering
				\includegraphics[width=\textwidth]{figures/KOcontrol08-nulltype=1-indepen-39.pdf}
			\end{minipage}%
		}
		\\
		\subfigure[$\pi_0 = 0.9$]{
			\begin{minipage}[t]{1\linewidth}
				\centering
				\includegraphics[width=\textwidth]{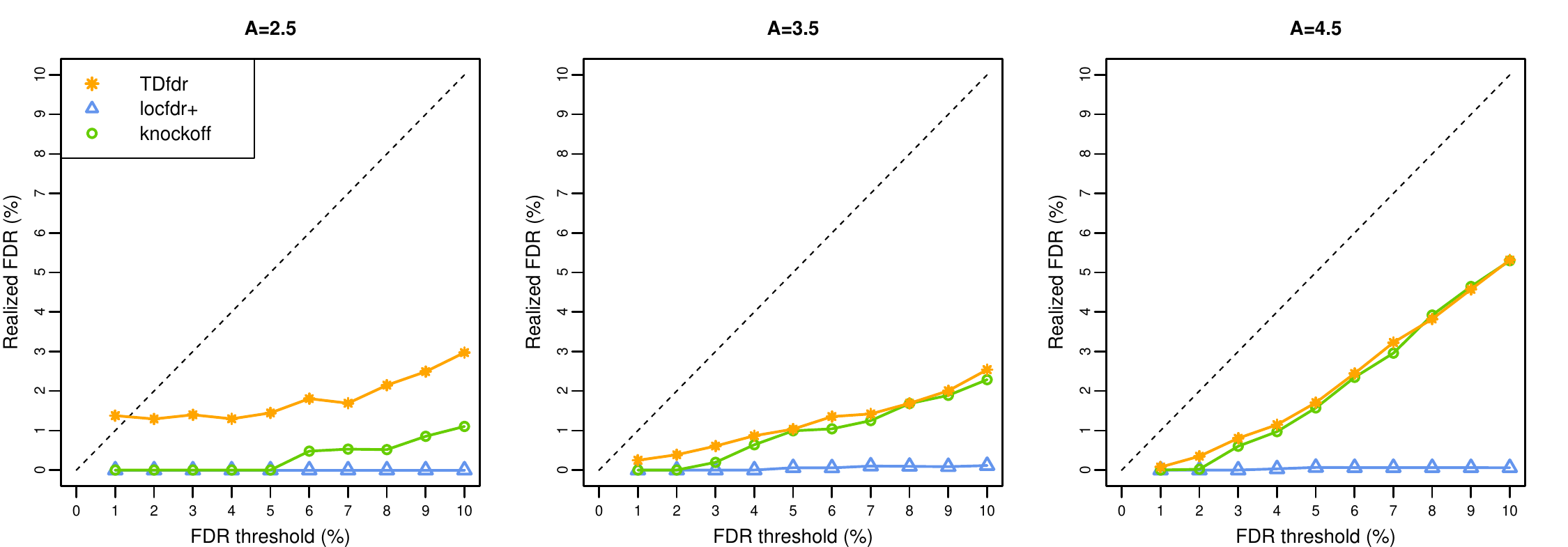}
			\end{minipage}%
		}
		\\
		\subfigure[$\pi_0 = 0.95$]{
			\begin{minipage}[t]{1\linewidth}
				\centering
				\includegraphics[width=\textwidth]{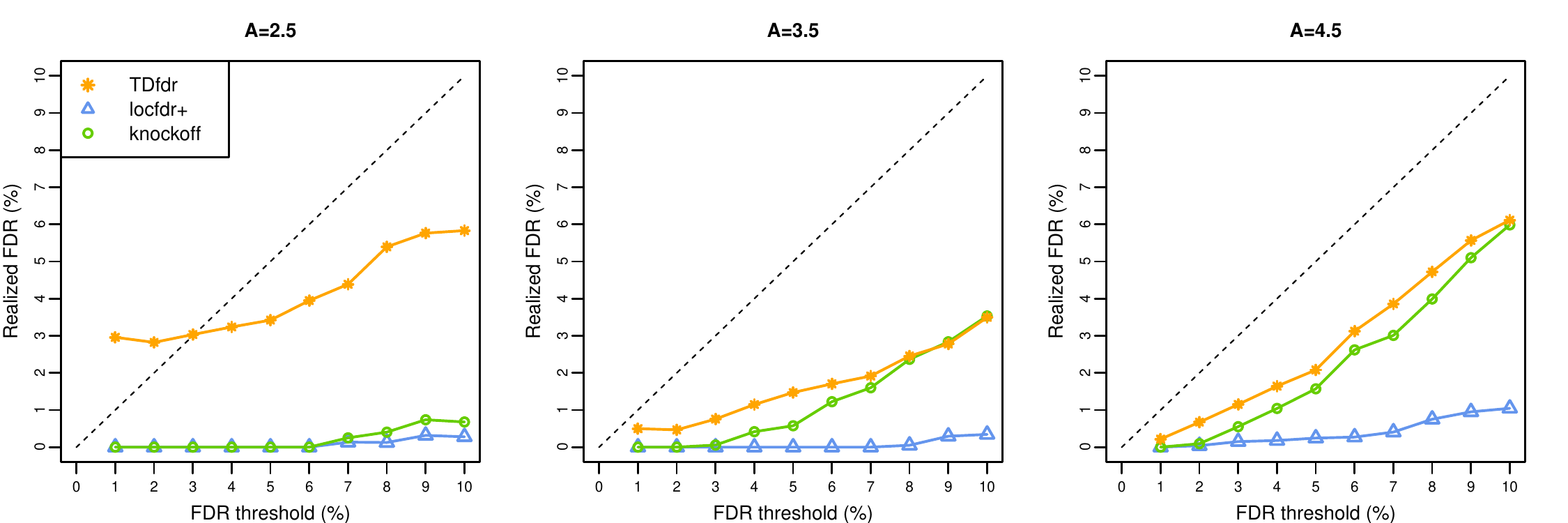}
			\end{minipage}%
		}%
		\centering
		\caption{FDR control results of regression data (independent cases)}
		\label{fig:koSimu-control}
	\end{figure}

	\begin{figure}[H]
		\centering
		\subfigure[$\pi_0 = 0.8$]{
			\begin{minipage}[t]{1\linewidth}
				\centering
				\includegraphics[width=\textwidth]{figures/KOcontrol08-nulltype=1-depen-39.pdf}
			\end{minipage}%
		}
		\\
		\subfigure[$\pi_0 = 0.9$]{
			\begin{minipage}[t]{1\linewidth}
				\centering
				\includegraphics[width=\textwidth]{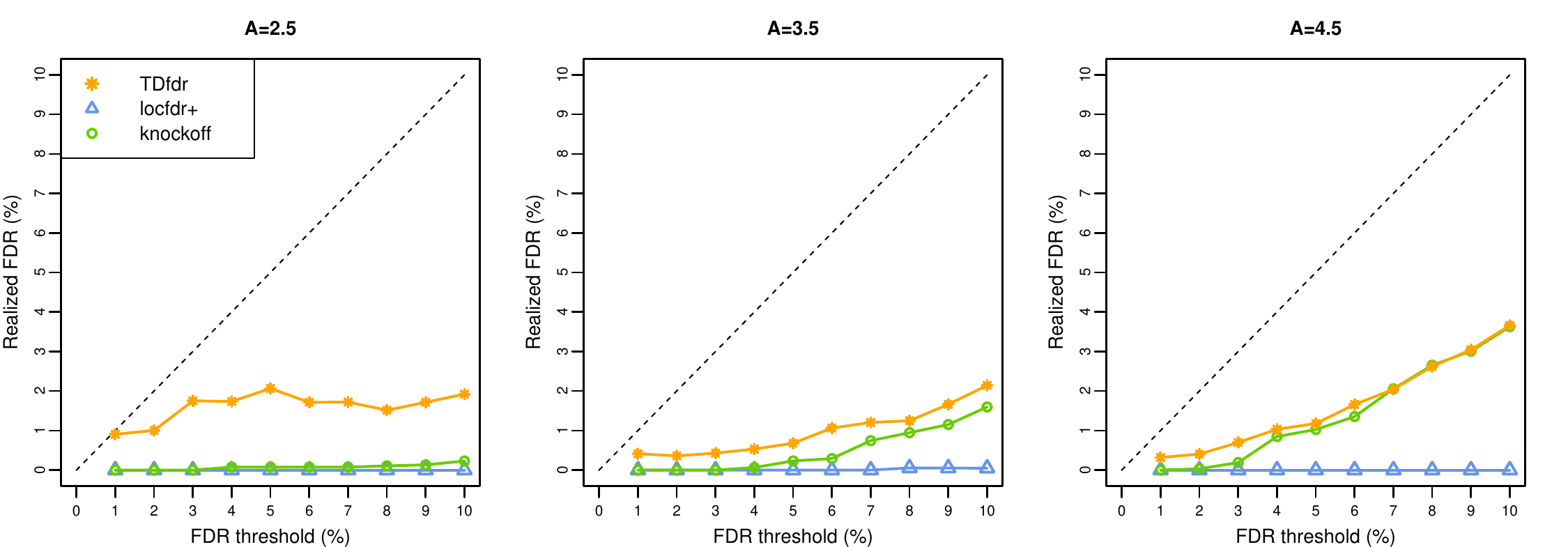}
			\end{minipage}%
		}%
		\\
		\subfigure[$\pi_0 = 0.95$]{
			\begin{minipage}[t]{1\linewidth}
				\centering
				\includegraphics[width=\textwidth]{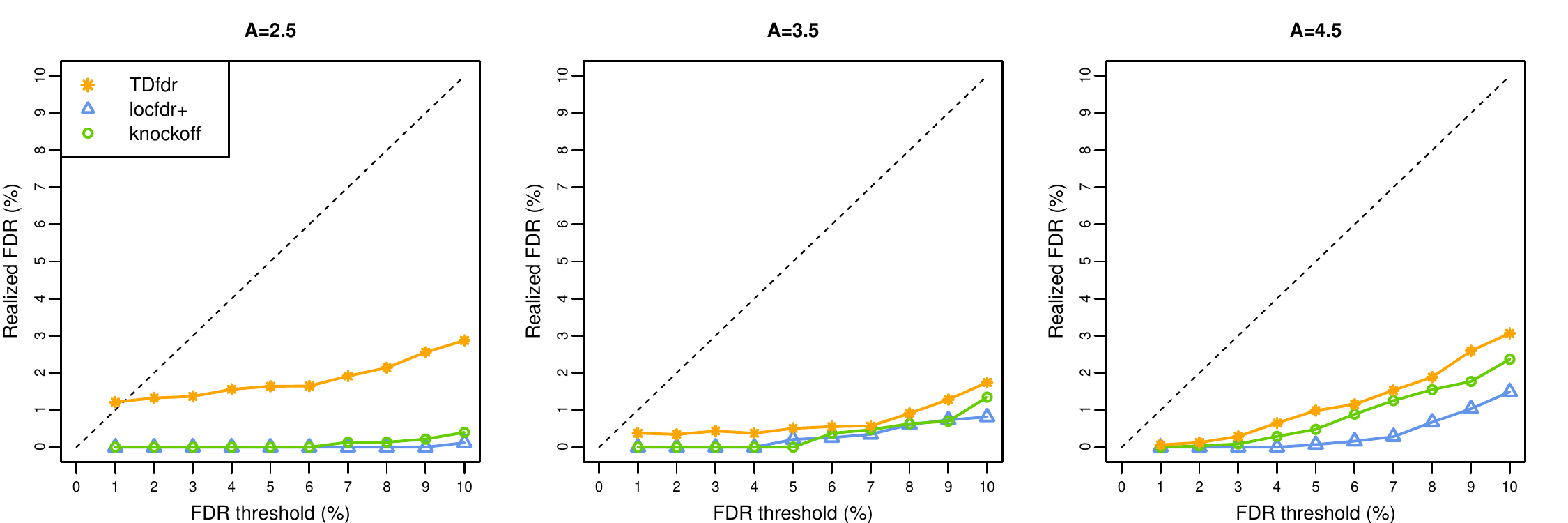}
			\end{minipage}%
		}%
		\centering
		\caption{FDR control results of regression data (dependent cases)}
		\label{fig:koSimu-control-corr}
	\end{figure}

	\begin{figure}[H]
		\centering
		\subfigure[$\pi_0 = 0.8$]{
			\begin{minipage}[t]{1\linewidth}
				\centering
				\includegraphics[width=\textwidth]{figures/KOpower08-nulltype=1-indepen-312.pdf}
			\end{minipage}%
		}%
		\\
		\subfigure[$\pi_0 = 0.9$]{
			\begin{minipage}[t]{1\linewidth}
				\centering
				\includegraphics[width=\textwidth]{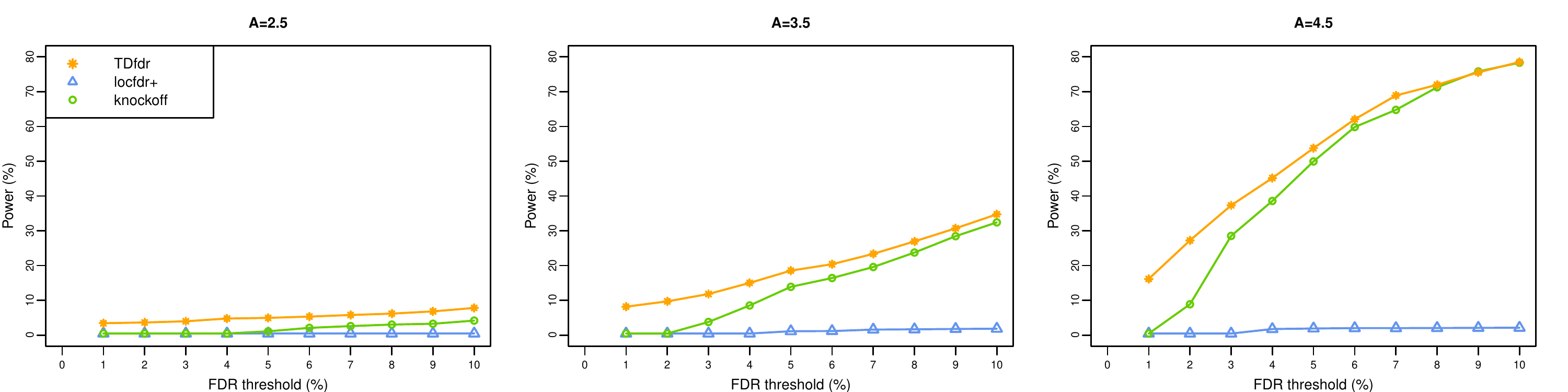}
			\end{minipage}%
		}%
		\\
		\subfigure[$\pi_0 = 0.95$]{
			\begin{minipage}[t]{1\linewidth}
				\centering
				\includegraphics[width=\textwidth]{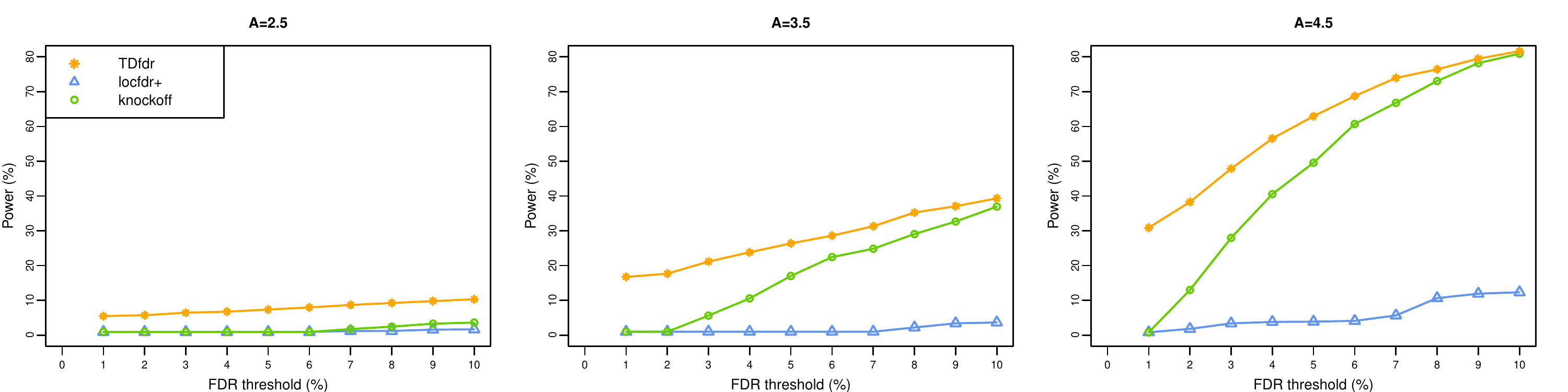}
			\end{minipage}%
		}%
		\centering
		\caption{Powers of regression data (independent cases)}
		\label{fig:koSimu-power}
	\end{figure}

	\begin{figure}[H]
		\centering
		\subfigure[$\pi_0 = 0.8$]{
			\begin{minipage}[t]{1\linewidth}
				\centering
				\includegraphics[width=\textwidth]{figures/KOpower08-nulltype=1-depen-312.pdf}
			\end{minipage}%
		}%
		\\
		\subfigure[$\pi_0 = 0.9$]{
			\begin{minipage}[t]{1\linewidth}
				\centering
				\includegraphics[width=\textwidth]{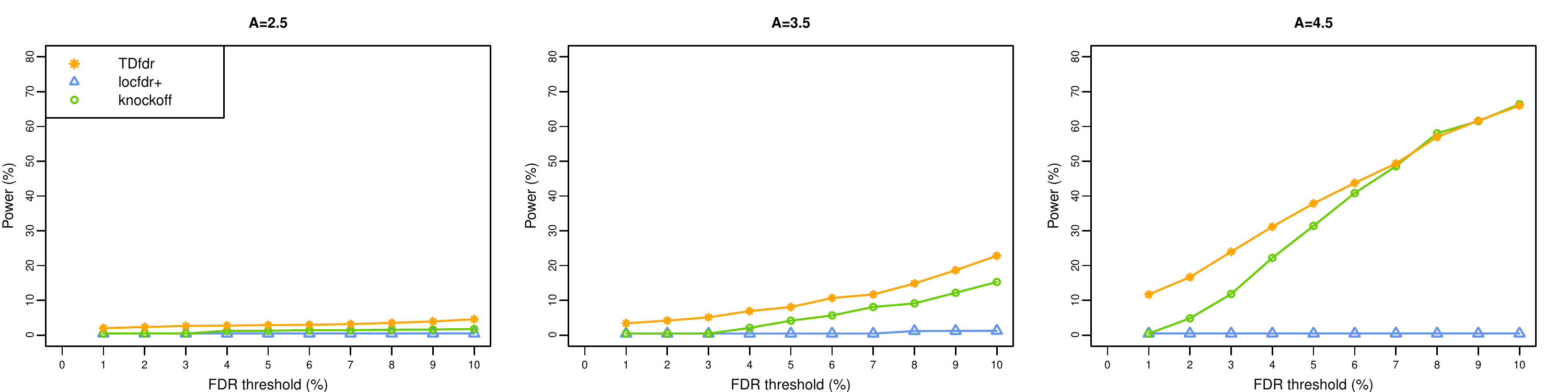}
			\end{minipage}%
		}%
		\\
		\subfigure[$\pi_0 = 0.95$]{
			\begin{minipage}[t]{1\linewidth}
				\centering
				\includegraphics[width=\textwidth]{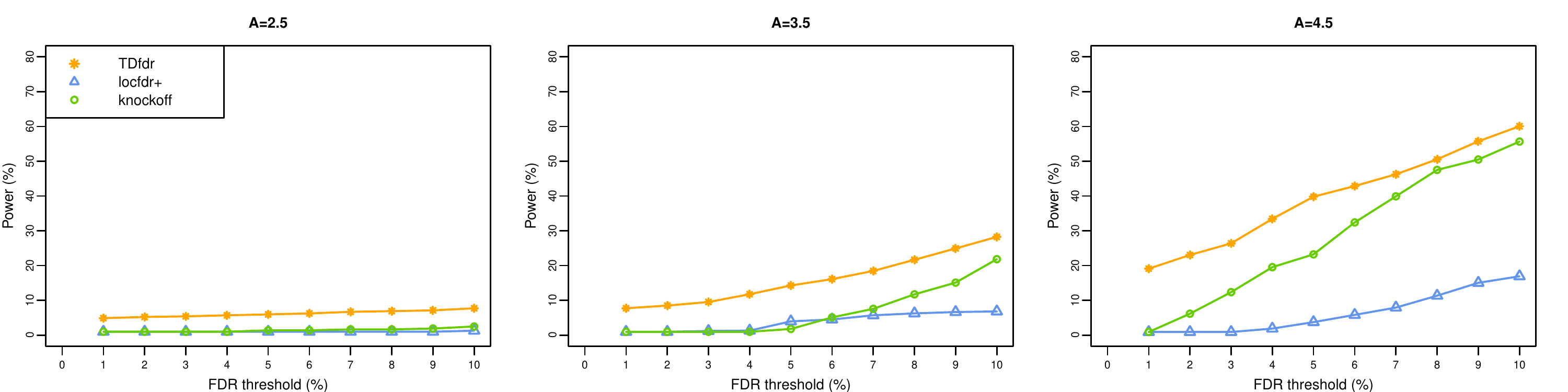}
			\end{minipage}%
		}%
		\centering
		\caption{Powers of regression data (dependent cases)}
		\label{fig:koSimu-power-corr}
	\end{figure}

	\newpage

	\subsection{COVID-19 data analysis results}\label{sec:S3}

	\begin{table}[H]
		\scriptsize
		\setlength{\abovecaptionskip}{0pt}%
		\setlength{\belowcaptionskip}{10pt}%
		\captionsetup{font={small}}
		\caption{Sample grouping details}
		\label{table:samplegroup}
		\begin{center}
			\scalebox{1}{
				\begin{threeparttable}
					\begin{tabular*}{0.7\textwidth}{ccl}
						\hline
						Group name & Group size & Description  \\
						\hline
						Severe  & 28 & Serum samples from severe COVID-19 patients \\
						Nonsevere  & 37 & Serum samples from nonsevere COVID-19 patients \\
						Non-COVID-19  & 25 & Serum samples from non-COVID-19$^{*}$ patients \\
						Healthy  & 28 & Serum samples from healthy subjects\\
						\hline
					\end{tabular*}
					\vspace{0pt}
					\begin{tablenotes}
						\scriptsize
						\item[*] Non-COVID-19 patients represent those who are negative for the SARS-CoV-2 nucleic acid test but have clinical characteristics similar to COVID-19 patients 
					\end{tablenotes}
			\end{threeparttable}}
		\end{center}
	\end{table}

	\begin{figure}[H]
		\centering
		\includegraphics[width=0.4\textwidth]{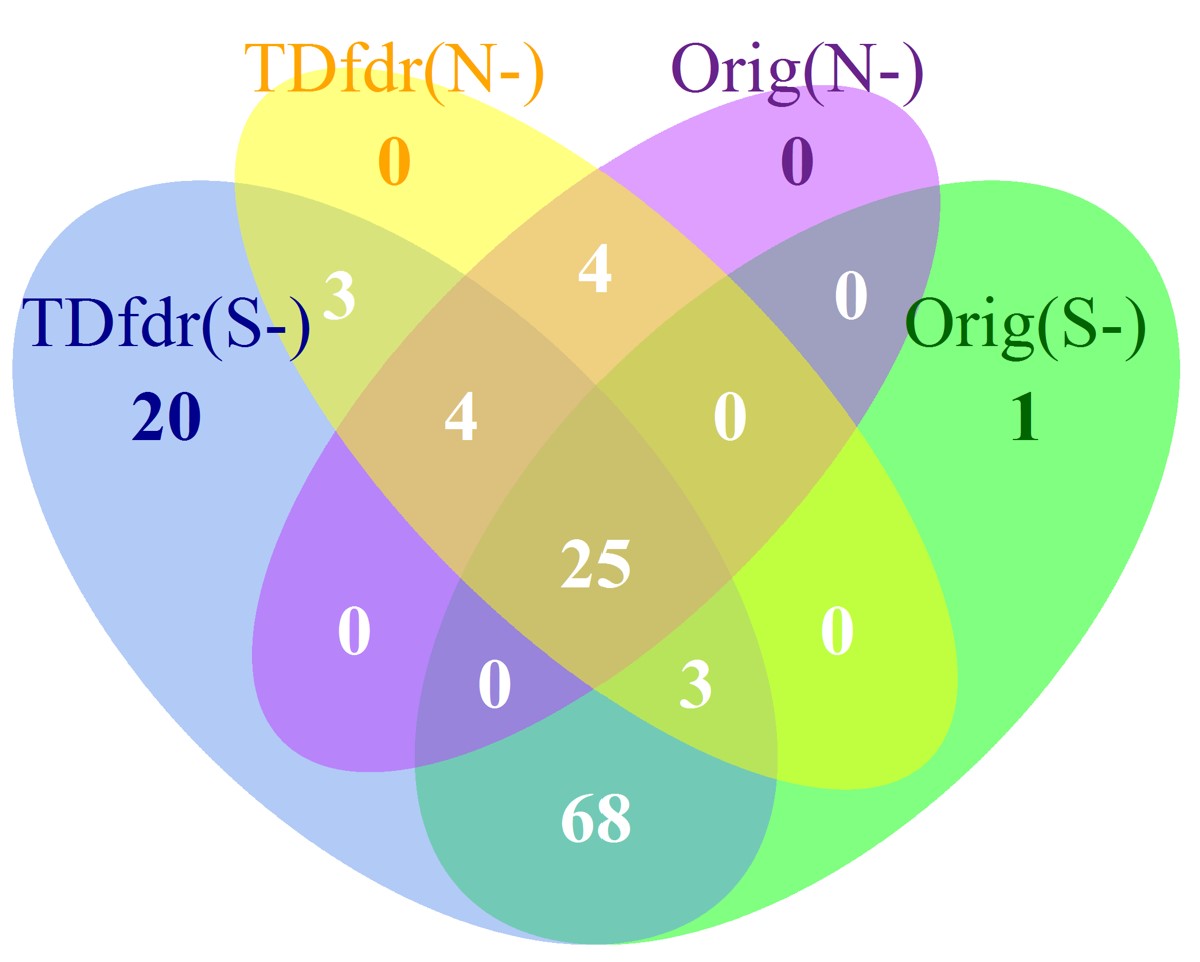}
		\caption{Comparison Venn plot of TDfdr and original results. "TDfdr(S-)" represents the results of set operation (Severe vs. Healthy) $\backslash$ (Non-COVID-19 vs. Healthy) from TDfdr, and "TDfdr(N-)" represents the results of set operation (Nonsevere vs. Healthy) $\backslash$ (Non-COVID-19 vs. Healthy) from TDfdr. Similarly, "Orig(S-)" and "Orig(N-)" represents the results of the corresponding set operations from the original study.}
		\label{fig:Venn}
	\end{figure}

	\begin{figure}[H]
		\centering
		\includegraphics[width=1\textwidth]{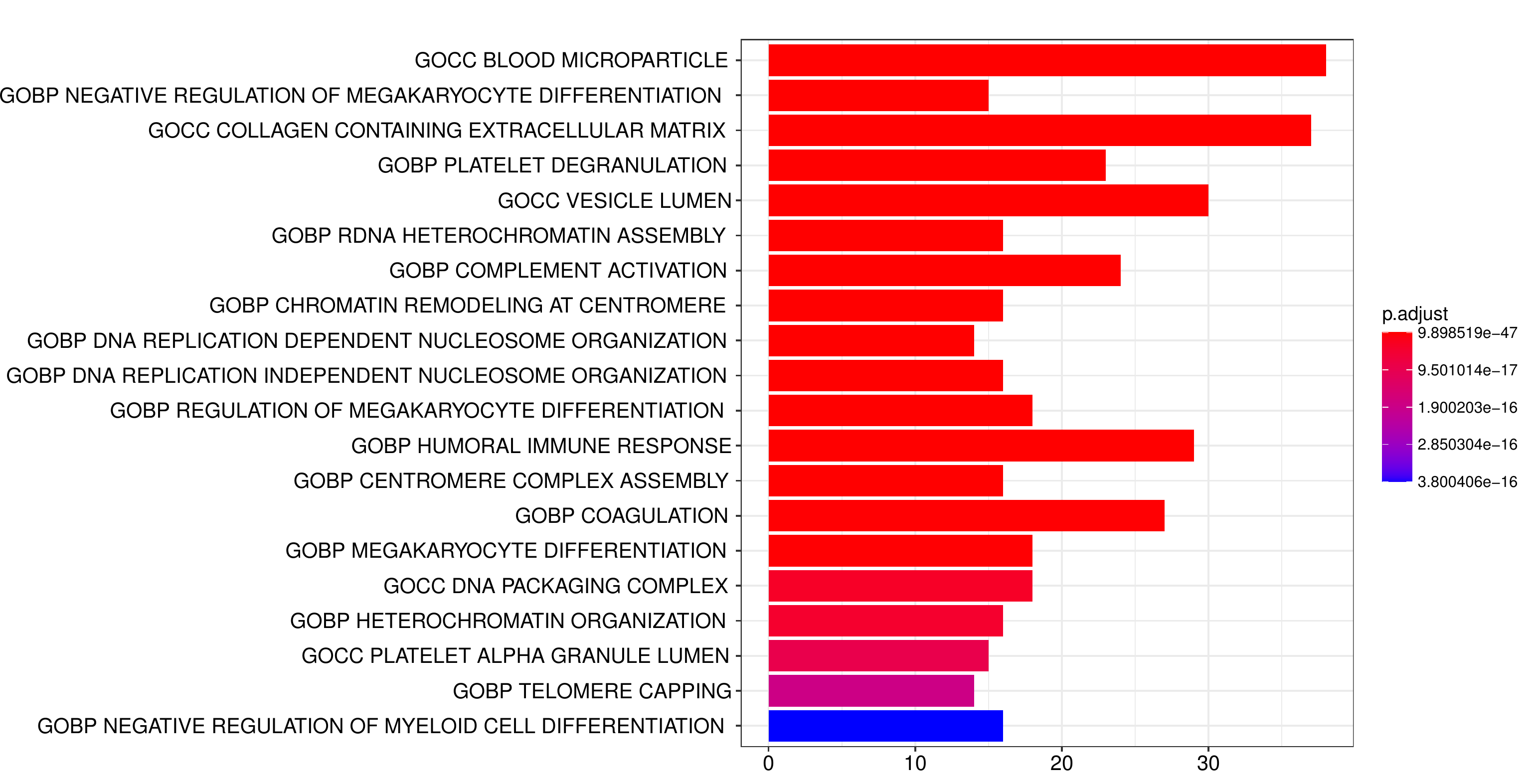}
		\caption{Pathways associated to the 123 severe proteins}
		\label{fig:pathway}
	\end{figure}

\end{document}